\pdfoutput=1
\documentclass[a4paper,fleqn,usenatbib]{mnras}

\usepackage[T1]{fontenc}
\usepackage{ae,aecompl}

\usepackage{graphicx}	
\usepackage{amsmath}	
\usepackage{amssymb}	
\usepackage{amsfonts}
\usepackage{pslatex}
\usepackage{color}
\usepackage{url}
\usepackage{float}
\usepackage[caption = false]{subfig}

\def\lesssim{\mathrel{\hbox{\rlap{\hbox{\lower3pt\hbox{$\sim$}}}\hbox{\raise2pt\hbox{$<$}}}}}
\def\gtrsim{\mathrel{\hbox{\rlap{\hbox{\lower3pt\hbox{$\sim$}}}\hbox{\raise2pt\hbox{$>$}}}}}

\newcommand{\lxt}{$L_{\rm X}-T$}

\title[The \lxt relation from XCS]{The \textit{XMM} Cluster Survey: joint modelling of the \lxt scaling relation for clusters and groups of galaxies}

\author[Ebrahimpour et al. (XCS collaboration)]{Leyla~Ebrahimpour,$^{1,2}$\thanks{E-mail: Leyla.Ebrahimpour@astro.up.pt} Pedro~T. P.~Viana,$^{1,2}$ Maria~Manolopoulou,$^{3}$
\newauthor Carlos~Vergara-Cervantes,$^{4}$ A.~Kathy~Romer,$^{4}$ Sunayana~Bhargava,$^{4}$ Paul~Giles,$^{4}$ 
\newauthor Alberto~Bermeo-Hernandez,$^{4}$ Chris~A.~Collins,$^{5}$ Matt~Hilton,$^{6}$ Ben~Hoyle,$^{7}$ 
\newauthor Andrew~R.~Liddle,$^{3,8}$ Robert~G.~Mann,$^{3}$ Julian~A.~Mayers,$^{4}$ Christopher~J.~Miller,$^{9}$  
\newauthor Robert~C.~Nichol,$^{10}$ Philip~J.~Rooney,$^{4}$ Martin~Sahl{\'e}n,$^{11}$ John~P.~Stott$^{12}$ 
\\
\\
$^{1}$~Instituto de Astrof\'{\i}sica e Ci\^{e}ncias do Espa\c{c}o, Universidade do Porto, CAUP, Rua das Estrelas, 4150-762 Porto, Portugal\\
$^{2}$~Departamento de F\'{\i}sica e Astronomia, Faculdade de Ci\^{e}ncias, Universidade do Porto, Rua do Campo Alegre, 687, 4169-007 Porto, Portugal\\
$^{3}$~Institute for Astronomy, University of Edinburgh, Royal Observatory, Blackford Hill, Edinburgh, EH9 3HJ, UK\\
$^{4}$~Astronomy Centre, University of Sussex, Falmer, Brighton, BN1 9QH, UK\\
$^{5}$~Astrophysics Research Institute, Liverpool John Moores University, IC2 Building, Liverpool Science Park, 146 Brownlow Hill, Liverpool L3 5RF, UK\\
$^{6}$~Astrophysics \& Cosmology Research Unit, School of Mathematics, Statistics \& Computer Science, University of KwaZulu-Natal, Westville Campus, Durban 4041, SA\\ 
$^{7}$~Universitaets-Sternwarte, Fakultaet fuer Physik, Ludwig-Maximilians Universitaet Muenchen, Scheinerstr. 1, D-81679 Muenchen, Germany\\
$^{8}$~Instituto de Astrof\'{\i}sica e Ci\^{e}ncias do Espa\c{c}o, Faculdade de Ci\^{e}ncias, Universidade de Lisboa, Edif\'{\i}cio C8, Campo Grande, 1769-016 Lisboa, Portugal\\
$^{9}$~Astronomy Department, University of Michigan, Ann Arbor, MI 48109, USA\\ 
$^{10}$~Institute of Cosmology and Gravitation, Dennis Sciama Building, Burnaby Road, Portsmouth, PO1 3FX, UK\\ 
$^{11}$~Department of Physics and Astronomy, Uppsala University, Box 516, SE-751 20 Uppsala, Sweden\\
$^{12}$~Department of Physics, Lancaster University, Lancaster LA1 4YB, UK
}

\date{Accepted XXX. Received YYY; in original form ZZZ}

\pubyear{2018}

\begin{document}

\label{firstpage}

\pagerange{\pageref{firstpage}--\pageref{lastpage}}

\maketitle

\begin{abstract}
We characterize the X-ray luminosity--temperature ($L_{\rm X}-T$) relation using a sample of 353 clusters and groups of galaxies with temperatures in excess of 1 keV, spanning the redshift range $0.1 < z < 0.6$, the largest ever assembled for this purpose. All systems are part of the \textit{XMM-Newton} Cluster Survey (XCS), and have also been independently identified in Sloan Digital Sky Survey (SDSS) data using the redMaPPer algorithm. We allow for redshift evolution of the normalisation and intrinsic scatter of the $L_{\rm X}-T$ relation, as well as, for the first time, the possibility of a temperature-dependent change-point in the exponent of such relation. However, we do not find strong statistical support for deviations from the usual modelling of the $L_{\rm X}-T$ relation as a single power-law, where the normalisation evolves self-similarly and the scatter remains constant with time. Nevertheless, assuming {\it a priori} the existence of the type of deviations considered, then faster evolution than the self-similar expectation for the normalisation of the $L_{\rm X}-T$ relation is favoured, as well as a decrease with redshift in the scatter about the $L_{\rm X}-T$ relation. Further, the preferred location for a change-point is then close to 2 keV, possibly marking the transition between the group and cluster regimes. Our results also indicate an increase in the power-law exponent of the $L_{\rm X}-T$ relation when moving from the group to the cluster regime, and faster evolution in the former with respect to the later, driving the temperature-dependent change-point towards higher values with redshift.
\end{abstract}


\section{Introduction}
\label{s_intro}

Clusters and groups of galaxies are the largest gravitationally bound structures in the universe. They are very complex systems, within which galaxies and the intracluster/intergroup medium (ICM/ICG) interact and evolve with time, in a background of gravitationally dominant dark matter. Clusters and groups can be detected and characterised in multiple regions of the electromagnetic spectrum: the ICM/ICG emits X-rays, mostly through the bremsstrahlung effect, while the Sunyaev-Zel\ensuremath{'}dovich effect produced by the ICM/ICG leaves an imprint in the microwave background radiation; their galaxy population is most clearly seen in the optical and infrared wavebands. Several of the quantities that can be estimated by observing clusters and groups in those regions of the electromagnetic spectrum, like the temperature and X-ray luminosity, the integrated comptonization parameter, and richness-related measures, seem to obey different scaling behaviours depending on whether one considers the cluster or the group regime \citep[e.g.][]{Sun_2012,Giodini_2013}. This suggests these structures have formed and/or evolved with time in distinct ways. Thus, the study of the different behaviours exhibited by cluster and group scaling relations can shed light on the disparate physical processes behind the assembly of those structures. It can also help define the distinction between the group and cluster regimes. This could prove to be particularly useful when it comes to using cluster properties to impose constraints on cosmological parameters \citep[e.g.][]{Vikhlinin_2009, MantzCosmo_2010, Rozo_2010, Allen_2011, BCC_2014, Mantz_2014, deHaan_2016, Planck_2016}. The reliability of these depends on the correct modelling of cluster scaling relations, used to link cluster mass to more directly observable cluster properties, as well as on the assembly of cluster samples devoid of objects, like groups, that may follow different scaling relations.

For example, it is expected that clusters and groups exhibit a different scaling relation between X-ray luminosity ($L_{\rm X}$) and temperature ($T$) in the context of the self-similar model \citep[e.g.][]{Sun_2012, Zou_2016}. Although the contrary has been recently claimed for observed clusters and groups \citep[][]{Bharadwaj_2015, Lovisari_2015, Zou_2016}, in none of these analysis the $L_{\rm X}-T$ relation in the cluster and group regimes was characterised simultaneously. In fact, every study of the $L_{\rm X}-T$ relation either made no distinction between clusters and groups, or ad hoc assumptions were used to define cluster-only or group-only samples. In particular, it has been common practice to assume that clusters have temperatures exceeding a value somewhere between 2 and 3 keV, contrary to groups. In either case, biases may have been introduced in the derivation of the behaviour of the $L_{\rm X}-T$ relation for the objects dominating the sample considered. This may be one of the reasons for some inconsistency among results obtained in the many studies already devoted to this subject, although inadequate or absent corrections for sample selection effects have surely also played a role. Namely, contradictory results have been obtained when using the behaviour of the $L_{\rm X}-T$ relation to test whether the self-similar model \citet{Kaiser_1986} provides a reasonable description of how clusters and groups have formed and evolved. Most noticeably, while the self-similar model predicts $L_{\rm X} \propto T^2$, observations suggest instead a steeper relation, $L_{\rm X} \propto T^{2-3}$ \citep[e.g. see review by][]{Giodini_2013}. A consensus has been even more difficult to reach with respect to the evolution of the cluster and group $L_{\rm X}-T$ relations. While the results of some analysis are consistent with the prediction of self-similarity \citep[e.g.][]{MantzScaling_2010, Maughan_2012}, whereby the normalisation of the $L_{\rm X}-T$ relation is positively correlated with redshift, $z$, being proportional to $E(z)=\sqrt{\Omega_{\rm m}(1+z)^3+\Omega_\Lambda}$, other studies have found evidence for either zero or even a negative correlation \citep[e.g.][]{Reichert_2011, Hilton_2012, Clerc_2014}. Given that the self-similar model assumes that the properties of the ICM/ICG are determined by the action of gravitational collapse alone, any departure from self-similarity would indicate that additional sources of energy have been heating the ICM/ICG. The most probable are the Active Galactic Nuclei (AGN) known to reside in the centres of most clusters, and in a significant fraction of groups \citep[e.g.][]{Gitti_2012}. 

The structure of this paper is as follows. We start by reviewing the characteristics of the XCS and SDSS data considered in the analysis. Next, we describe the method used to characterise the $L_{\rm X}-T$ relation in the cluster and group regimes simultaneously. We then present the results obtained, discuss them, and present our conclusions.

We assume throughout the universe to be geometrically flat, with fractional densities of matter and dark energy (as a cosmological constant) to be $\Omega_{\rm m}=0.27$ and $\Omega_\Lambda=0.73$, respectively, with the Hubble constant being $H_0=70$ km s$^{-1}$ Mpc$^{-1}$.

\section{Data}
\label{s_data}

The \textit{XMM} Cluster Survey \citep[XCS\footnotemark\footnotetext{\url{http://www.xcs-home.org}};][]{Romer_2001} is a serendipitous search for galaxy clusters and groups in the \textit{XMM-Newton} Science Archive (XSA\footnotemark\footnotetext{\url{http://nxsa.esac.esa.int}}). The X-ray analysis methodology for the survey was originally described in \citet{LloydDavies_2011}, and then updated in \citet{Manolopoulou_2018}. In the Appendix, we use the sample considered in \citet{Hilton_2012} to illustrate the impact on the estimation of temperatures and X-ray luminosities of the changes described in \citet{Manolopoulou_2018}, as well as on the characterization of the $L_{\rm X}-T$ relation.

The first data release \citep[XCS-DR1;][]{Mehrtens_2011} contained a total of 401 X-ray selected clusters and groups with temperature and redshift estimates. Since then, we have been expanding the XCS, by analysing newer XSA data and using more recent optical cluster/group catalogues, namely those that are based on Sloan Digital Sky Survey \citep[SDSS;][]{York_2000} and Dark Energy Survey \citep[DES;][]{Flaugher_2005} data, to confirm our X-ray candidates. In \citet{Manolopoulou_2018} we present XCS-DR2-SDSS, the subset of the XCS Data Release 2 within the SDSS DR13 footprint \citep[][]{Albareti_2017}, comprising more than one thousand optically-confirmed groups and clusters, about two hundred of which are new discoveries. However, here we concentrate on a sample of groups and clusters that includes only 353 of those objects, which were also independently detected by applying the redMaPPer algorithm to SDSS-DR8 \citep[][]{Rykoff_2014}. We decided to consider only such objects in order to make the underlying selection function of our sample as simple as possible, just the result of the convolution of the XCS selection function with that for the SDSS-DR8-redMaPPer catalogue. Note that more objects than the 353 considered appear both in the XCS and in the SDSS-DR8-redMaPPer catalogue. The smaller size of our sample results mainly from retaining only the objects with a coefficient of variation for the temperature lower than $0.25$, thus ensuring that our temperature estimates are reliable \citep[for further details, see][]{Manolopoulou_2018}. We also restricted our sample to groups and clusters with a redshift equal or greater than 0.1, in order to avoid having to deal with the more complicated nature of the XCS selection function at lower redshifts, and with measured temperatures in excess of 1 keV, given that the XCS selection function has not been characterised for lower temperatures \citep{Sahlen_2009}.

The XCS identification (ID), SDSS-DR8-redMaPPer redshift \citep[][]{Rykoff_2014}, most probable values (i.e. modes) for the temperature and X-ray luminosity, as well as the associated 68\% confidence intervals derived through our X-ray analysis pipeline, for all 353 sample systems, are listed in a supplementary table available with the online edition of this article. The first 10 rows are shown in Table \ref{table_data}. 

\begin{table}
\caption{The first 10 rows of a supplementary table available with the online edition of this article, providing the XCS ID, SDSS-DR8-redMaPPer redshift \citep[][]{Rykoff_2014}, $z$, most probable values for the temperature, $T$ (in units of keV), and X-ray bolometric luminosity within $R_{500}$, $L_X$ (in units of $10^{44}$ erg s$^{-1}$), as well as the associated 68\% confidence intervals derived through our X-ray analysis pipeline, for all 353 sample systems.}
\label{table_data}
\def\arraystretch{1.5}
\begin{tabular}{|c|c|c|c|}
\hline
XCS ID & z & $T$ & $L_X$\\
\hline
${\rm XMMXCS\;J140101.9+025238.3}$ & $0.26$ & $6.63\,_{-0.02}^{+0.03}$ & $55.39\,_{-0.09}^{+0.09}$\\
${\rm XMMXCS\;J142601.0+374937.0}$ & $0.17$ & $8.65\,_{-0.10}^{+0.10}$ & $31.88\,_{-0.20}^{+0.20}$\\
${\rm XMMXCS\;J172009.8+263725.8}$ & $0.17$ & $6.15\,_{-0.04}^{+0.04}$ & $21.50\,_{-0.08}^{+0.08}$\\  
${\rm XMMXCS\;J224321.4-093550.2}$ & $0.44$ & $7.22\,_{-0.07}^{+0.09}$ & $33.11\,_{-0.09}^{+0.08}$\\
${\rm XMMXCS\;J212939.7+000516.9}$ & $0.25$ & $5.29\,_{-0.05}^{+0.05}$ & $22.74\,_{-0.12}^{+0.13}$\\ 
${\rm XMMXCS\;J075124.2+173057.7}$ & $0.19$ & $3.28\,_{-0.05}^{+0.05}$ & $2.86\,_{-0.02}^{+0.02}$\\  
${\rm XMMXCS\;J172227.0+320758.0}$ & $0.23$ & $7.21\,_{-0.09}^{+0.09}$ & $27.58\,_{-0.18}^{+0.17}$\\
${\rm XMMXCS\;J104044.2+395711.1}$ & $0.14$ & $3.74\,_{-0.04}^{+0.04}$ & $7.39\,_{-0.04}^{+0.04}$\\  
${\rm XMMXCS\;J111253.4+132640.2}$ & $0.17$ & $4.89\,_{-0.06}^{+0.06}$ & $6.12 \,_{-0.04}^{+0.04}$\\ 
${\rm XMMXCS\;J024803.3-033143.4}$  & $0.19$ & $3.76\,_{-0.04}^{+0.04}$ & $9.64\,_{-0.06}^{+0.06}$\\  
\hline
\end{tabular}
\end{table}

\begin{figure}
\includegraphics[width=8.0cm]{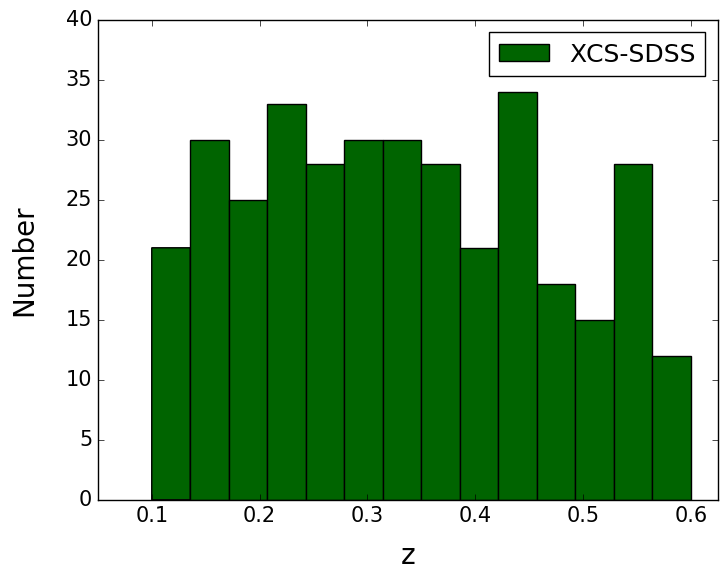}
\includegraphics[width=8.0cm]{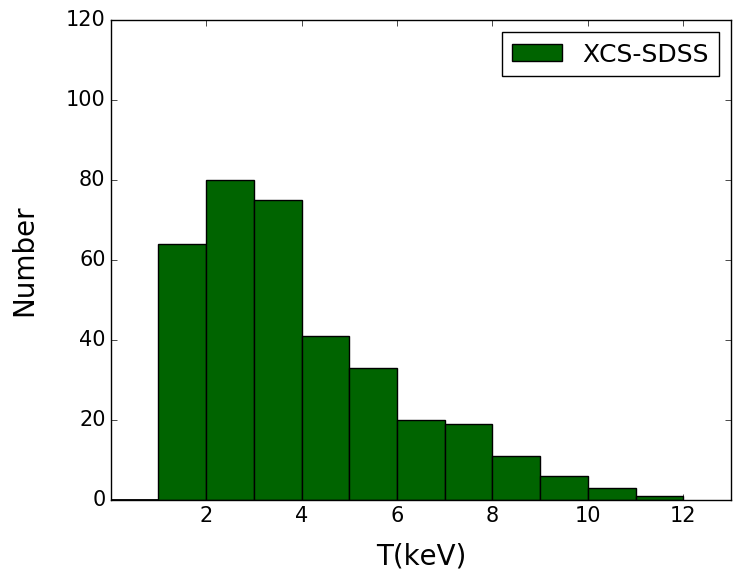}
\includegraphics[width=8.0cm]{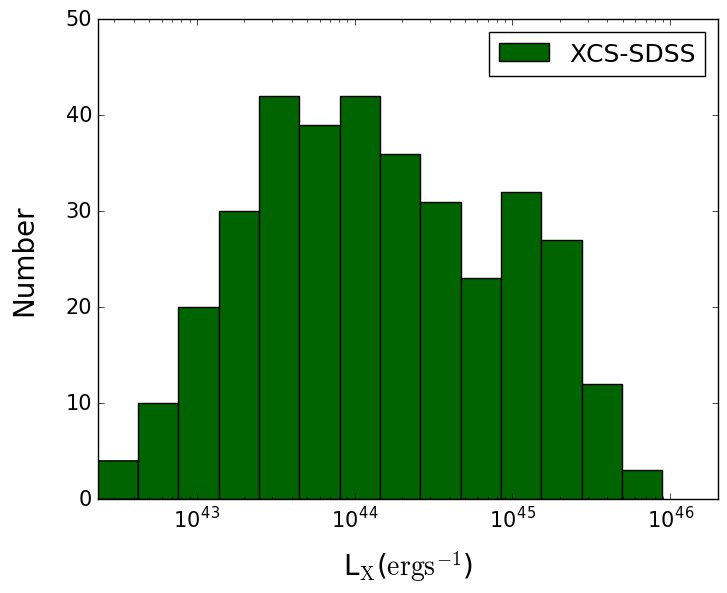}
\caption{The distributions of most probable values for the redshift, $z$, temperature, $T$ (in keV), and  X-ray luminosity, $L_X$ (in erg/s), given the data, for the 353 XCS-DR2-SDSS-DR8-redMaPPer groups and clusters used in this work.}
\label{zLT_histograms}
\end{figure}

\begin{figure}
\includegraphics[width=8.2cm]{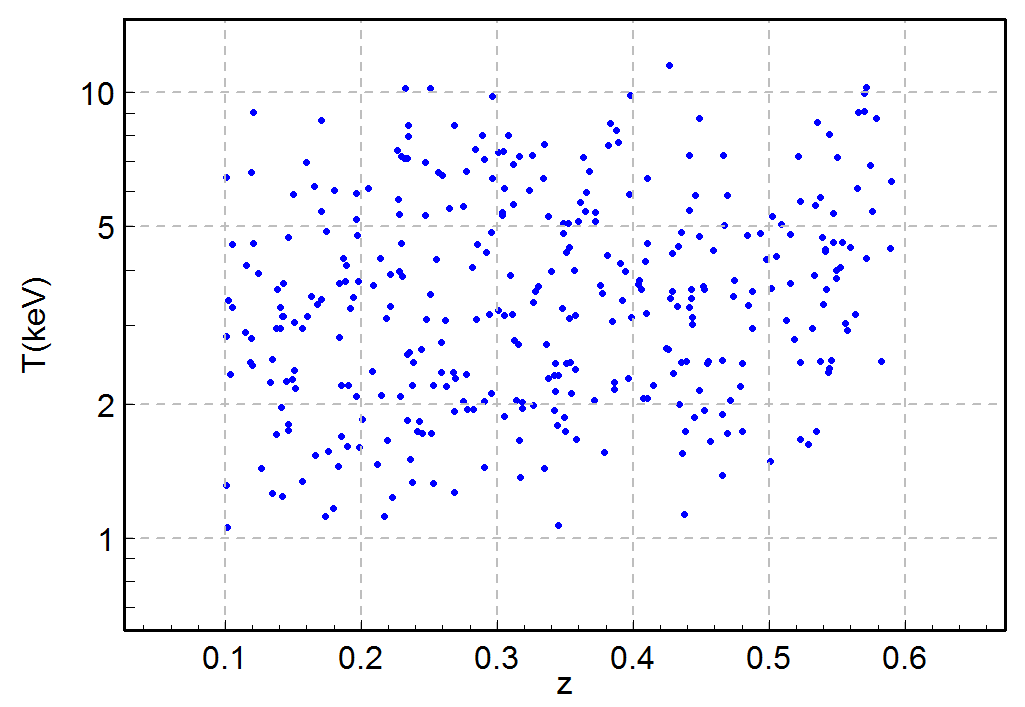}
\includegraphics[width=8.2cm]{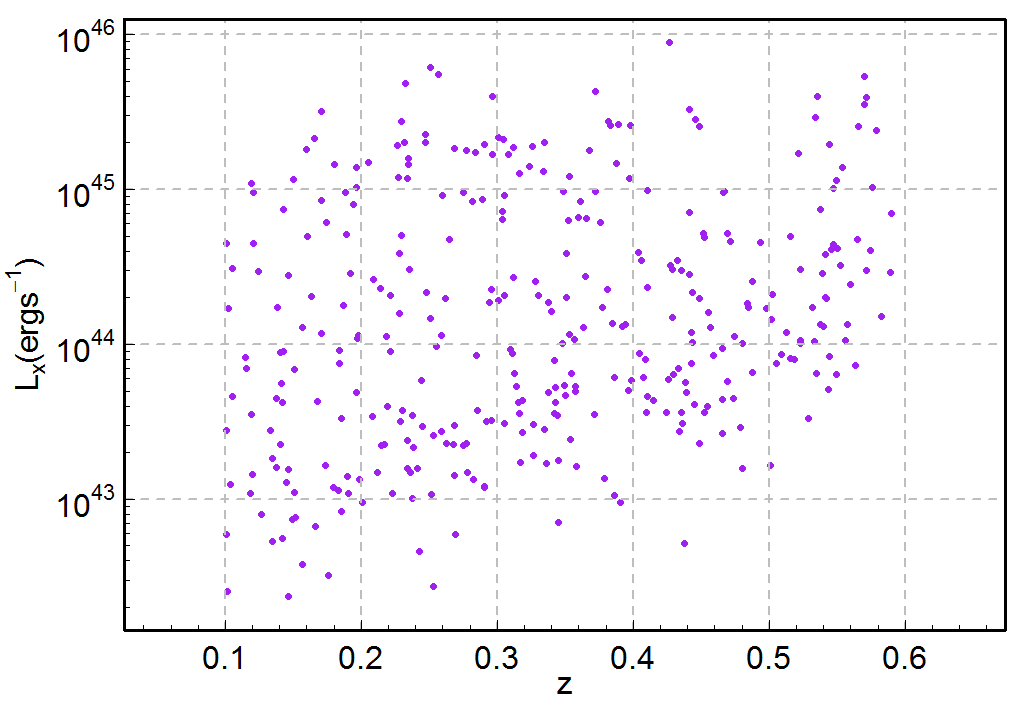}
\caption{Most probable values for temperature, $T$ (in keV, upper panel), and X-ray luminosity, $L_X$ (in erg/s, lower panel), versus redshift, given the data, for the 353 XCS-DR2-SDSS-DR8-redMaPPer groups and clusters used in this work.}
\label{zLT_distributions}
\end{figure}

The sample distributions of most probable values for the redshift, temperature and X-ray luminosity, given the data, are represented in Figs.~\ref{zLT_histograms} and ~\ref{zLT_distributions}. The groups and clusters span the redshift range $0.10$ -- $0.59$ (median $0.33$), temperature range $1.06$ -- $11.53$ keV (median $3.39$\,keV), and luminosity range $2.36\times10^{42}$ -- 8.90$\times10^{45}$ erg s$^{-1}$ (median $1.19\times10^{44}$ erg s$^{-1}$).

\section{Analysis}
\label{s_analysis}
\subsection{Bayesian framework}
\label{subs_bayes}
Cluster and group scaling relations have been most often modelled as a single power-law with a time-evolving normalization, following the self-similar expectation \citep[][]{Kaiser_1986}. Under such assumptions, the general form of the relation between two properties, e.g., the dependent (response), $Y$, and the independent (covariate), $X$, variables is
\begin{equation}
Y=\alpha\,X^{\beta}\,{F_{\rm z}}^{\gamma},
\label{eq1}
\end{equation}
which in logarithmic scale, is expressed as a linear relation,
\begin{equation}
\log (Y) = \alpha + \beta\log (X) + \gamma\log (F_{\rm z}) + \epsilon,
\label{eq2}
\end{equation}
where the factor depending on $F_{\rm z}$ allows for evolution with redshift, and $\epsilon$ represents the intrinsic scatter in log($Y$) about the regression relationship. We will assume the later to be described by a Gaussian distribution with mean zero and {\it a priori} unknown dispersion $\sigma$. This may be allowed to evolve with time according to
\begin{equation}
\sigma(z)= \sigma_0\,{F_{\rm z}}^{\gamma_\sigma}\,
\label{eq3}
\end{equation}
where $ \sigma_0$ is the zero redshift dispersion, and $\gamma_\sigma$ accounts for the time-evolution of the dispersion. 

Our interest lies in modelling the scaling relation between X-ray luminosity, $L_{X}$, and temperature, $T$. Therefore, we will re-write equation (\ref{eq2}) as
\begin{equation}
\log [L_{\rm X}/E(z)] = \alpha + \beta\log (T/5) + \gamma\log E(z) + \epsilon,
\label{eq4}
\end{equation}
where the logarithms are base 10, $L_{\rm X}$ is the bolometric X-ray luminosity and $T$ is the temperature, in units of erg s$^{-1}$ and keV respectively. Both are estimated iteratively using only data that falls within $R_{500}$, the radius enclosing an average density 500 times larger than the critical density $3H^2/8\pi G$. Note that $L_{\rm X}$ and $T$ are estimated using all data that falls within $R_{500}$, i.e. they are not core-excised quantities \citep[for further details, see][]{Manolopoulou_2018}. Since the luminosity has been scaled by $E^{-1}(z)$ in equation (\ref{eq3}), the self-similar model predicts $\gamma=0$, as well as $\beta=2$ \citep[][]{Kaiser_1986}. We set $F_{\rm z}$ equal to $E(z)$, instead of $(1+z)$, to enable a more straightforward comparison with the expectation for self-similar evolution when the exponent for the $L_{X}-T$ relation deviates from the self-similar prediction \citep{Maughan_2014}. The pivot temperature is fixed at 5\,keV, close to the mean of the sample (3.4 keV), to allow for comparison of our results with those of \citet[][]{Hilton_2012}. It is assumed that $z$ is not affected by measurement uncertainties, i.e. its actual value is perfectly known, given that typically $\sigma_z\sim0.01(1+z)$ for the clusters and groups in our sample \citep[][]{Rozo_2014,Rykoff_2014}. 

However, because the properties of clusters and groups may obey different scaling relations, we will also consider the possible existence of an abrupt transition or change-point in the scaling relation between $T$ and $L_{\rm X}$. This is achieved by introducing two extra parameters, $T_{\rm break}$ and $\beta_{\rm break}$ in our generic model, where the later takes the role of $\beta$ for $T < T_{\rm break}$. All other generic model parameters discussed so far, i.e. $\alpha$, $\gamma$, $\sigma$ and $\gamma_{\sigma}$, are assumed to be universal across the range of $ L_{\rm X}$ and $T$ values that we will consider, in order to ensure continuity between the cluster and group scaling regimes.

We will be using a Bayesian hierarchical modelling framework to constrain $T_{\rm i}$ and $L_{{\rm X},{\rm i}}$, for each $i$ cluster or group in our sample, as well as the hyper-parameters associated with each particular model we will consider, conditioned on all the X-ray and redshift data, $D\equiv\{D_{\rm i}\}$, we have gathered regarding those objects \cite[for a similar approach see e.g.][]{MantzScaling_2010,Maughan_2014}. This will be achieved through the characterization of the marginal posterior probability distributions for all $T_{\rm i}$ and $L_{{\rm X},{\rm i}}$, as well as for the hyper-parameters, in the context of each of those particular models. All these, numbered from 0 to 7, are nested inside the generic model (labelled as 7). Each is uniquely identified through its associated set of hyper-parameters, although models 0 and 1 have variants, denoted 0u and 1u, which we will later differentiate from the main model: 

\vspace{0.4cm}

\noindent$\theta_0\equiv(\alpha,\beta,\sigma)$

\noindent$\theta_1\equiv(\alpha,\beta,\sigma,\gamma)$

\noindent$\theta_2\equiv(\alpha,\beta,\sigma,\gamma_{\sigma})$

\noindent$\theta_3\equiv(\alpha,\beta,\sigma,T_{\rm break},\beta_{\rm break})$

\noindent$\theta_4\equiv(\alpha,\beta,\sigma,\gamma,\gamma_{\sigma})$

\noindent$\theta_5\equiv(\alpha,\beta,\sigma,\gamma,T_{\rm break},\beta_{\rm break})$

\noindent$\theta_6\equiv(\alpha,\beta,\sigma,\gamma_{\sigma},T_{\rm break},\beta_{\rm break})$

\noindent$\theta_7\equiv(\alpha,\beta,\sigma,\gamma,\gamma_{\sigma},T_{\rm break},\beta_{\rm break})$\,.

\vspace{0.4cm}

In practice, it is computationally easier to work with the (generic) linear model represented by equation (\ref{eq4}), where $X_{\rm i}\equiv \log(T_{\rm i}/5)$ and $Y_{\rm i}\equiv \log[L_{{\rm X},{\rm i}}/E(z_{\rm i})]$. Therefore, we will first characterize the marginal posterior probability distributions for each $X_{\rm i}$ and $Y_{\rm i}$, and then transform them into marginal posterior probability distributions for each $T_{\rm i}$ and $L_{{\rm X},{\rm i}}$. Marginal distributions are obtained by integrating the joint posterior probability distribution, which can be evaluated, up to a normalization constant, through the expressions
\begin{align}
P(\{X_{\rm i}, Y_{\rm i}\}, \theta | D)&\propto\!P(D | \{X_{\rm i}, Y_{\rm i}\}, \theta)P(\{X_{\rm i}, Y_{\rm i}\}, \theta)\\
&\propto\!P(\{D_{\rm i}\} | \{X_{\rm i}, Y_{\rm i}\}, \theta)P(\{X_{\rm i}, Y_{\rm i}\} | \theta)P(\theta)\\
&\propto\!\prod_{\rm i=1}^{n} P(D_{\rm i} | X_{\rm i}, Y_{\rm i}, \theta)P(Y_{\rm i} | X_{\rm i}, \theta)P(X_{\rm i} | \theta)P(\theta),
\label{eq6}
\end{align}

\noindent where $n$ is the number of objects in our sample. We assumed, as usual, that the data pertaining to each cluster or group, $D_{\rm i}$, does not hold information about the properties of any other object. In the last expression, the first term is known as the likelihood, and in our case it will be independent of the values that any particular set of hyper-parameters, generically labelled $\theta$, may take. The product of the second and third terms is the prior probability distribution for each set of hidden (sometimes also called latent) variables, $X_{\rm i}$ and $ Y_{\rm i}$, conditioned on the hyper-parameters. We assumed all such sets to be mutually independent. The second term is just the Gaussian distribution that describes the stochastic relation between $X$ and $Y$, characterized through equation (\ref{eq2}), and thus is in fact independent of the object $i$ under consideration. The last term in the expression is the joint prior probability distribution for the hyper-parameters (also known as hyper-prior). 

The R package for Bayesian {\bf LI}near {\bf R}egression in {\bf A}stronomy ({\bf LIRA}), described in \cite{LIRA}, and available both at CRAN\footnotemark\footnotetext{\url{https://cran.r-project.org/web/packages/lira/}} and GitHub\footnotemark,\footnotetext{\url{https://github.com/msereno/lira/}} was used to efficiently explore, using Gibbs sampling, the joint posterior probability distribution. The Monte Carlo Markov Chains (MCMC) thus assembled was then used to characterize the marginal posterior distributions for each set of the hidden variables $T_{\rm i}$ and $L_{{\rm X},{\rm i}}$, as well as for the hyper-parameters, $\theta$. We assessed chain convergence using the Gelman-Rubin diagnostic \citep[][]{Gelman_1992}, applied independently to each hidden variable and hyper-parameter of the model under consideration.

The likelihood, $P(D_{\rm i} | X_{\rm i}, Y_{\rm i})$, is affected by the presence of sample selection effects. In our case, these change the relative abundance of objects in the sample with respect to what is found in the population, as a function of both the covariate (X) and response (Y). In the former case, this comes about from our selection of only objects with a measured temperature in excess of 1 keV, while in the later case it results from the XCS selection function \citep{Sahlen_2009}, which we model as a bivariate sigmoid function and favours the detection of objects with higher X-ray luminosity at any given redshift or temperature. The impact of both sample selection effects on the likelihood is modelled using the capabilities LIRA has for such purpose \citep[][]{LIRA,Sereno_2017}. The raw likelihood (i.e. in the absence of sample selection effects) with respect to each set of the hidden variables $T_{\rm i}$ and $L_{{\rm X},{\rm i}}$ is assumed to take the form of a bivariate split Gaussian probability distribution, with the co-variance terms in all quadrants set to zero. Such a distribution is equivalent to the product of two independent univariate split Gaussian distributions \citep[][]{Villani_2006}, which is the type of distribution used to approximate the results of our X-ray analysis pipeline \citep{LloydDavies_2011,Manolopoulou_2018}, based on XSPEC\footnotemark\footnotetext{\url{https://heasarc.gsfc.nasa.gov/docs/xanadu/xspec/}}. We then follow \citet[][]{Sereno_2015} and use expressions (27) and (28) provided by \citet[][]{DAgostini_2000} to estimate the expected values and terms of the covariance matrix associated with the bivariate Gaussian distribution we assume describes the likelihood with respect to each set of the hidden variables, $X_{\rm i}$ and $Y_{\rm i}$, in the absence of sample selection effects.

The prior probability distribution for each hidden variable $X_{\rm i}$, conditioned on the hyper-parameters, $\theta$, i.e. $P(X_{\rm i} | \theta)$, is assumed to be well described by a time-evolving mixture of Gaussian distributions \citep{LIRA}. This allows the analysis to take into account: (1) the effect of the underlying temperature function, which describes the relative abundance of groups and clusters in the overall population as a function of their temperature; (2) the SDSS-DR8-redMaPPer selection function, which make the incorporation of an object in our sample less probable as its temperature decreases and redshift increases. We explored the cases where the number of Gaussians in the mixture was fixed {\it a priori} to either one or two. However, in the case of a mixture of two Gaussians we were unable to ascertain convergence of the MCMC for all the hyper-parameters associated with the more complex models. Therefore, we chose instead to restrict the mixture to just one Gaussian. Nevertheless, the marginal posterior distributions thus obtained are very similar to those that result from assuming a mixture of two Gaussians. Every particular set of hyper-parameters is henceforth expanded to include: $\mu$, the Gaussian (mixture) mean; $\sigma_\mu$, the Gaussian (mixture) dispersion; and $\gamma_\mu$, the parameter that allows for time-evolution of the Gaussian (mixture) mean. Note that LIRA can allow for a possible evolution of $\sigma_\mu$ with time, but we did not consider {\it a priori} the possibility of it being different from zero. We are not directly interested in characterizing the marginal probability distributions for the three Gaussian (mixture) hyper-parameters, and therefore from here on we will omit any explicit reference to them, except when discussing the results.

All hyper-parameters are assumed to be independent {\it a priori}, and thus for every particular model the hyper-prior is just the product of the prior probability distributions for each hyper-parameter associated with the model. We set these to the default choices in LIRA \citep{LIRA}, adapted to the characteristics of our models and sample. Namely, uniform prior distributions in the ranges $[40, 50]$ for $\alpha$, $[-0.7, 0.3]$ for $\log (T_{\rm break}/5)$, and $[-2.0, 1.0]$ for $\mu$, were assumed, in order to bracket, within a conservative margin, the sample values for $\log [L_{\rm X}/E(z)]$, in the first case, and $\log (T/5)$, in later two cases. We checked a posteriori that changing these limits did not have a significant impact on the results. Further, the prior distributions for $\beta$, $\beta_{\rm break}$, $\gamma$, $\gamma_\sigma$ and $\gamma_\mu$, were set to Student's t distributions with one degree of freedom (also known as standard Cauchy distributions). This choice ensures that the results are invariant under a rotation of the coordinate system used to identify the location of the sample objects, i.e. it ensures that all angles the possible regression lines make with, for example, the horizontal coordinate axis are equally probable {\it a priori} \citep[see e.g.][]{Toussaint_2011,Sharma_2017}. Finally, a $\Gamma$ prior distribution, with shape and rate parameters equal to 0.0001, was assumed for both $1/\sigma^2$ and $1/\sigma_\mu^2$. This choice makes the prior distribution with respect to the variances, $\sigma^2$ and $\sigma_\mu^2$, nearly scale-invariant.

\subsection{Model comparison}
\label{subs_model}

There are several approaches to the issue of model comparison, even within Bayesian statistics. The most prominent ones attempt either to identify the true underlying model or to find the most useful model. Useful in this context is usually considered to mean the ability of a given model to make predictions about future or yet undisclosed observations. In the first approach, the true underlying model is assumed {\it a priori} to be within the (finite) set of models considered in the analysis, contrary to the second approach. This has led, for example \citet{Vehtari_2012}, to propose labelling those two contrasting views of model comparison as $M$-closed and $M$-open, respectively. 

Model comparison usually proceeds through the evaluation of the probability of each model given the data, also known as the marginal likelihood or evidence, in the $M$-closed approach, or using cross-validation in the $M$-open approach \cite[e.g.][]{Piironen_2017}. However, both evidence estimation and cross-validation are computationally expensive to perform, which has led to several proposed approximations [see e.g. section 2.5 of \citet{Sharma_2017} for a short review]. There is a growing consensus that the Widely applicable (or Watanabe) Bayesian Information Criterion (WBIC) and the Widely Applicable (or Watanabe-Akaike) Information Criterion (WAIC) are among the most accurate and theoretically justified methods to approximate the results of model comparison based on, respectively, evidence estimation and cross-validation \cite[][]{Gelman_2014,Friel_2017,Piironen_2017,Sharma_2017,Vehtari_2017}.

In our case, WBIC is given by \cite[][]{Watanabe_2013,Sharma_2017}

\begin{align}
{\rm WBIC}&={\mathbb E}^{\rm 1/\ln n}\left[-\ln P(D | \{X_{\rm i}, Y_{\rm i}\}, \theta)\right]\\
&=\tfrac{\int-\ln P(D | \{X_{\rm i}, Y_{\rm i}\}, \theta)[P(D | \{X_{\rm i}, Y_{\rm i}\}, \theta)]^{\rm 1/\ln n}P(\{X_{\rm i}, Y_{\rm i}\}, \theta)dX_{\rm i}dY_{\rm i}d\theta}{\int[P(D | \{X_{\rm i}, Y_{\rm i}\}, \theta)]^{\rm 1/\ln n}P(\{X_{\rm i}, Y_{\rm i}\}, \theta)dX_{\rm i}dY_{\rm i}d\theta}
\label{eq7}
\end{align}

\noindent where $n$ is the number of independent data points, i.e. $n=353$ for the sample under consideration. The WBIC is an approximation to the Bayes free energy, i.e. $-\ln P(D|M_\theta)$, where $P(D|M_\theta)$ is the evidence associated to a particular model $M_\theta$, with hyper-parameters $\theta$. The difference between the expectation values for WBIC and the Bayes free energy is $O(\ln\ln n)$ \cite[][]{Watanabe_2013}. Comparing WBIC values associated with different particular models is thus equivalent to comparing their posterior probability distributions under the assumption of equality for their prior probability distributions, which under no reason to assume otherwise, is what we will be doing. 

The expectation value for some parameter-dependent function can be readily estimated using the output of a MCMC used to characterize the joint posterior probability distribution of the parameter(s) the function depends upon. This is achieved by averaging over the values such function takes given the values of the parameter(s) at each point in the MCMC. In our case, this means that

\begin{equation}
{\rm WBIC}\simeq-\frac{1}{m}\sum_{\rm j=1}^{m}\sum_{\rm i=1}^{n}\ln P(D_{\rm i} | X_{\rm i,j}, Y_{\rm i,j}, \theta_{\rm j})
\label{eq8}
\end{equation}

\noindent where $j$ identifies a point in a given MCMC of length $m$ used to characterize the joint posterior probability distribution that is proportional to $[P(D | \{X_{\rm i}, Y_{\rm i}\}, \theta)]^{\rm 1/\ln n}P(\{X_{\rm i}, Y_{\rm i}\}, \theta)$. Given that the summations in the MCMC approximation to WBIC can be interchanged, its estimation can be thought as resulting from the sum of the $n$ independent components ${\rm WBIC}_i\equiv-\frac{1}{m}\sum_{\rm j=1}^{m}\ln P(D_{\rm i} | X_{\rm i,j}, Y_{\rm i,j}, \theta_{\rm j})$. Hence, the product of $n$ by their variance equals the sample variance associated with the WBIC estimate \cite[][]{Vehtari_2017}.

On the other hand, WAIC is given by \cite[][]{Watanabe_2010,Piironen_2017} 

\begin{equation}
{\rm WAIC}=T+V/n
\label{eq9}
\end{equation}

\noindent where, in our case,

\begin{equation}
T=\sum_{\rm i=1}^{n}T_i
\label{eq10}
\end{equation}

\noindent is the training utility, loss or error, with

\begin{equation}
T_i\equiv-\frac{1}{n}\ln{\mathbb E}\left[P(D_{\rm i} | X_{\rm i}, Y_{\rm i}, \theta)\right]\,,
\label{eq11}
\end{equation}

\noindent and

\begin{equation}
V=\sum_{\rm i=1}^{n}V_i
\label{eq12}
\end{equation}

\noindent is the functional variance, with

\begin{equation}
V_i\equiv{\mathbb E}\left[(\ln P(D_{\rm i} | X_{\rm i}, Y_{\rm i}, \theta))^2\right]-\{{\mathbb E}[\ln P(D_{\rm i} | X_{\rm i}, Y_{\rm i}, \theta)]\}^2\,.
\label{eq13}
\end{equation}

\noindent In these expressions, the expectations are calculated with respect to the joint posterior probability distribution of $\{X_{\rm i}, Y_{\rm i}\}$ and $\theta$. The WAIC is asymptotically equal to the result of Bayesian leave-one-out cross-validation (LOO-CV), as well as to the generalization utility or loss, with the difference between their expectation values being $O(1/n^2)$ \cite[][]{Watanabe_2010,Watanabe_2013}. In our case, the estimation of WAIC can be also thought as the sum of the $n$ independent components ${\rm WAIC}_i\equiv T_i+V_i/n$. Therefore, the product of $n$ by their variance equals the sample variance associated with the WAIC estimate \cite[][]{Vehtari_2017}.

The expectations that need to be calculated to estimate each component ${\rm WAIC}_i$ can be obtained using MCMC output, in a similar way to what was done with regards to the estimation of WBIC. Therefore, in our case

\begin{equation}
T_i=\ln\left[\frac{1}{m}\sum_{\rm j=1}^{m}P(D_{\rm i} | X_{\rm i,j}, Y_{\rm i,j}, \theta_{\rm j})\right]
\label{eq14}
\end{equation}

\noindent and

\begin{footnotesize}
\begin{equation}
V_i=\frac{1}{m}\sum_{\rm j=1}^{m}\left[\ln P(D_{\rm i} | X_{\rm i,j}, Y_{\rm i,j}, \theta_{\rm j})\right]^2-\left[\frac{1}{m}\sum_{\rm j=1}^{m}\ln P(D_{\rm i} | X_{\rm i,j}, Y_{\rm i,j}, \theta_{\rm j})\right]^2
\label{eq15}
\end{equation}
\end{footnotesize}

\noindent where again $j$ identifies a point in a given MCMC of length $m$, but now used to characterize the joint posterior probability distribution that is proportional to $P(D | \{X_{\rm i}, Y_{\rm i}\}, \theta)P(\{X_{\rm i}, Y_{\rm i}\}, \theta)$. 

\section{Results}
\label{s_results}

\begin{figure}
\includegraphics[width=\columnwidth]{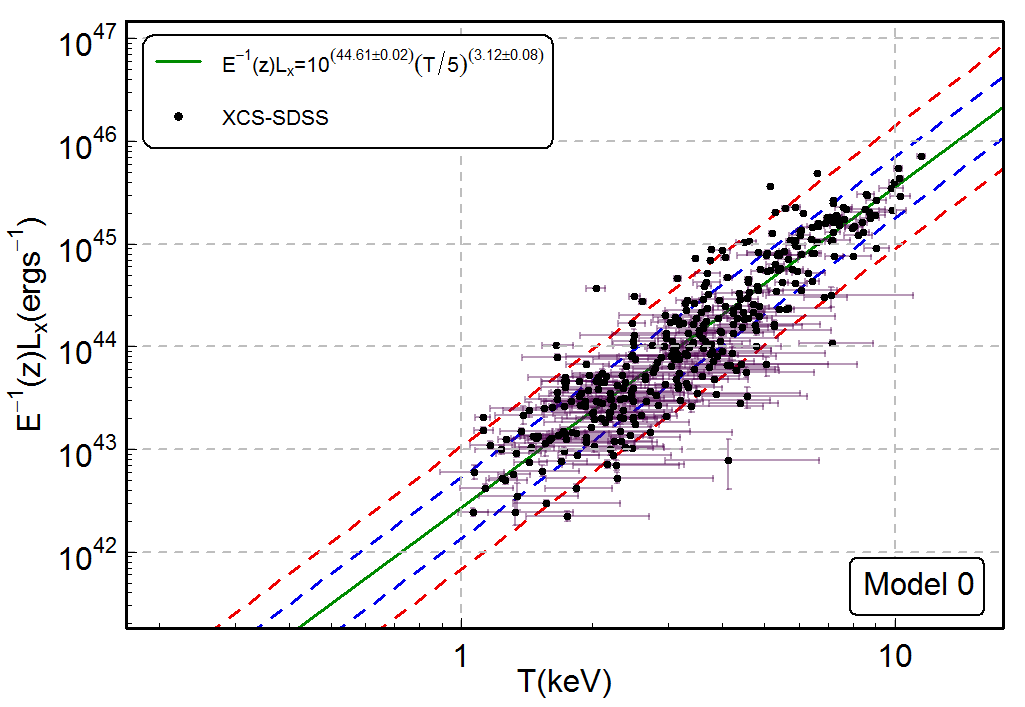}
\includegraphics[width=\columnwidth]{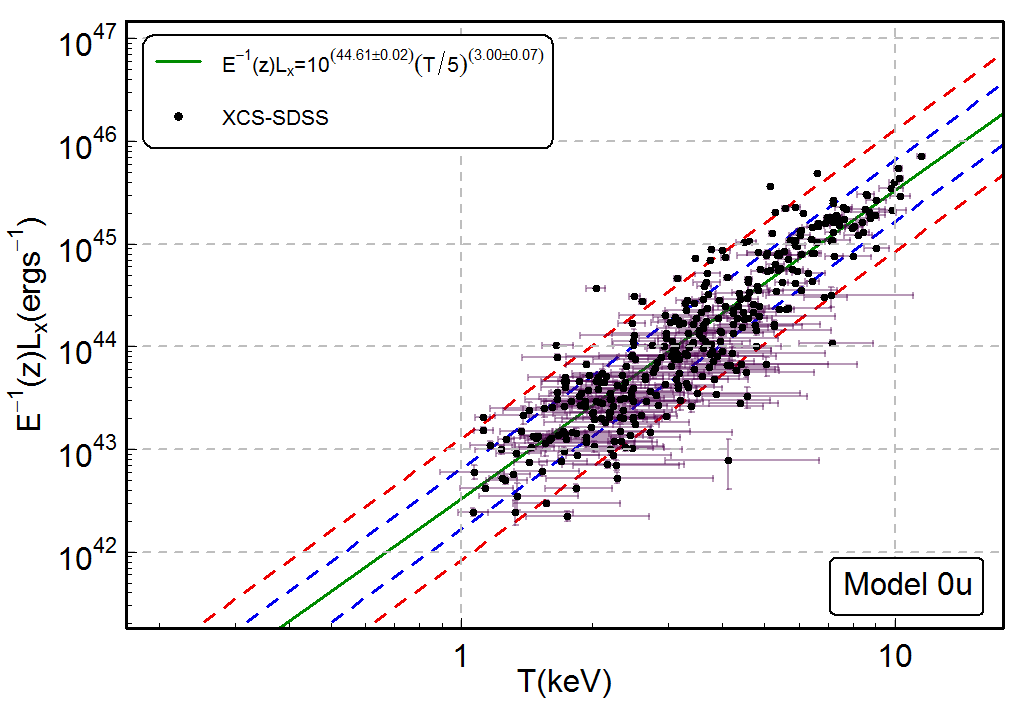}
\caption{The \lxt relation assuming the simplest model, with (top, model 0) and without (bottom, model 0u) sample selection effects taken into account. Each point indicates the most probable values for the temperature, $T$, in keV, and re-scaled bolometric X-ray luminosity, $ L_{\rm X}$, with respect to each of the 353 systems in our sample, given the X-ray data. The self-similar re-scaling of $L_{\rm X}$ with the cluster redshift, $z$, is performed dividing $L_{\rm X}$ by $E(z)$. The error bars associated with each point identify the $1\sigma$ uncertainty intervals. The solid regression line was determined by fixing the model hyper-parameters to their expected (i.e. mean) values. The amplitude of the intrinsic vertical scatter about the regression line is indicated by the dashed lines ($1\sigma$, inner blue; $2\sigma$, outer red).}
\label{fig_model_0}
\end{figure}

Let us start by considering the simplest model, with hyper-parameters $\theta_0$, where only the normalisation, $\alpha$, slope, $\beta$, and scatter, $\sigma$, of the $\log L_{\rm X}$ - $\log T$ relation are free to vary within the respective prior distributions. The time-evolution parameters $\gamma$ and $\gamma_\sigma$ are set to zero, and there is no change-point. In the upper panel of Fig. \ref{fig_model_0}, the regression line obtained assuming the expected (i.e. mean) values for the relevant model hyper-parameters, can be compared with the distribution of data points. In order to allow for comparison with previous results, we also show (lower panel) the regression line obtained when the XCS selection function and the imposition of a lower threshold of 1 keV in observed temperature for inclusion in the sample are not considered in the modelling, and the prior distribution for the temperature is (implicitly) assumed to be uniform. This version of model 0, we label as 0u, hence does not take into account any type of sample selection effect.

\begin{figure}
\includegraphics[width=\columnwidth]{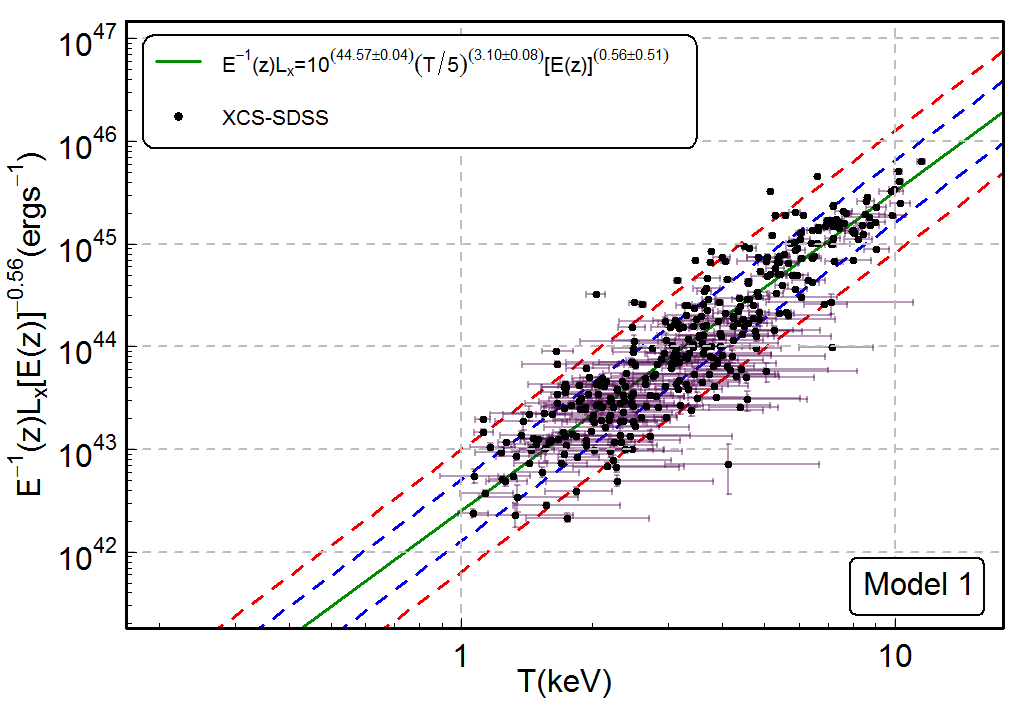}
\includegraphics[width=\columnwidth]{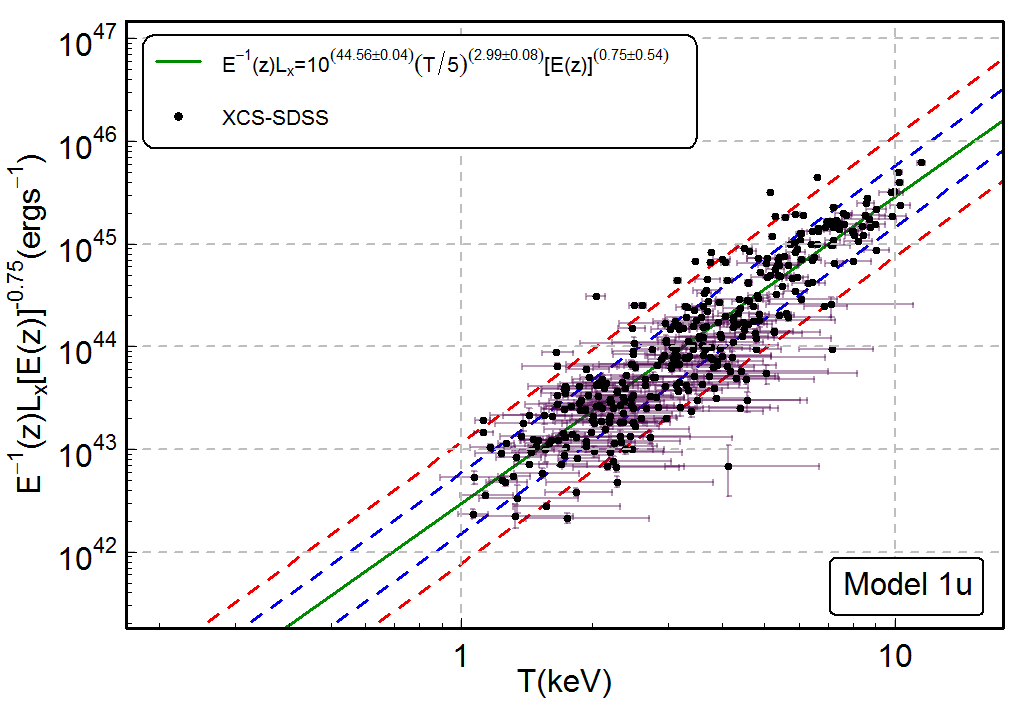}
\caption{The \lxt relation assuming model 1, with (top, model 1) and without (bottom, model 1u) sample selection effects taken into account. In this case, the re-scaling of $L_{\rm X}$ with the cluster redshift, $z$, is performed dividing $L_{\rm X}$ by $E(z)^{1+\overline\gamma}$, where $\overline\gamma$ is the expected (i.e. mean) value for $\gamma$.}
\label{fig_model_1}
\end{figure}

\begin{figure}
\includegraphics[width=\columnwidth]{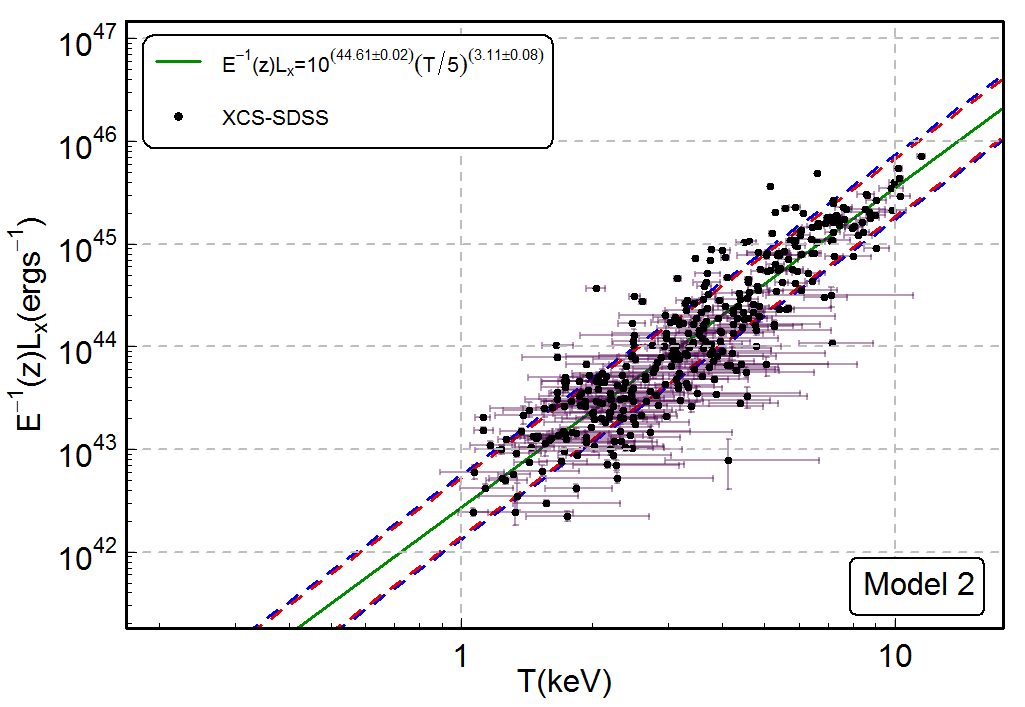}
\caption{The \lxt relation assuming model 2. The prior distribution for the temperature is modelled through a time-evolving Gaussian. The dashed lines now indicate the amplitude of the intrinsic vertical scatter about the regression line only at the $1\sigma$ level, for the minimum (0.1, outer blue) and maximum (0.6, inner red) sample redshifts.}
\label{fig_model_2}
\end{figure}

\begin{figure}
\includegraphics[width=\columnwidth]{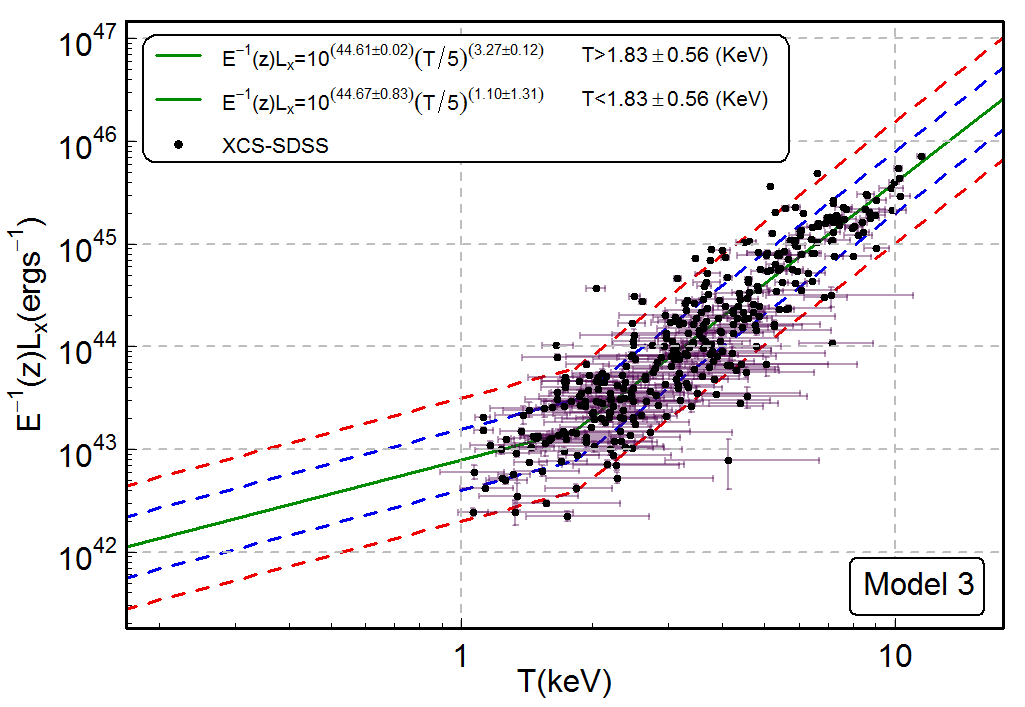}
\caption{The \lxt relation assuming model 3. The prior distribution for the temperature is modelled through a time-evolving Gaussian.}
\label{fig_model_3}
\end{figure}

\begin{figure*}
\includegraphics[width=17.2cm]{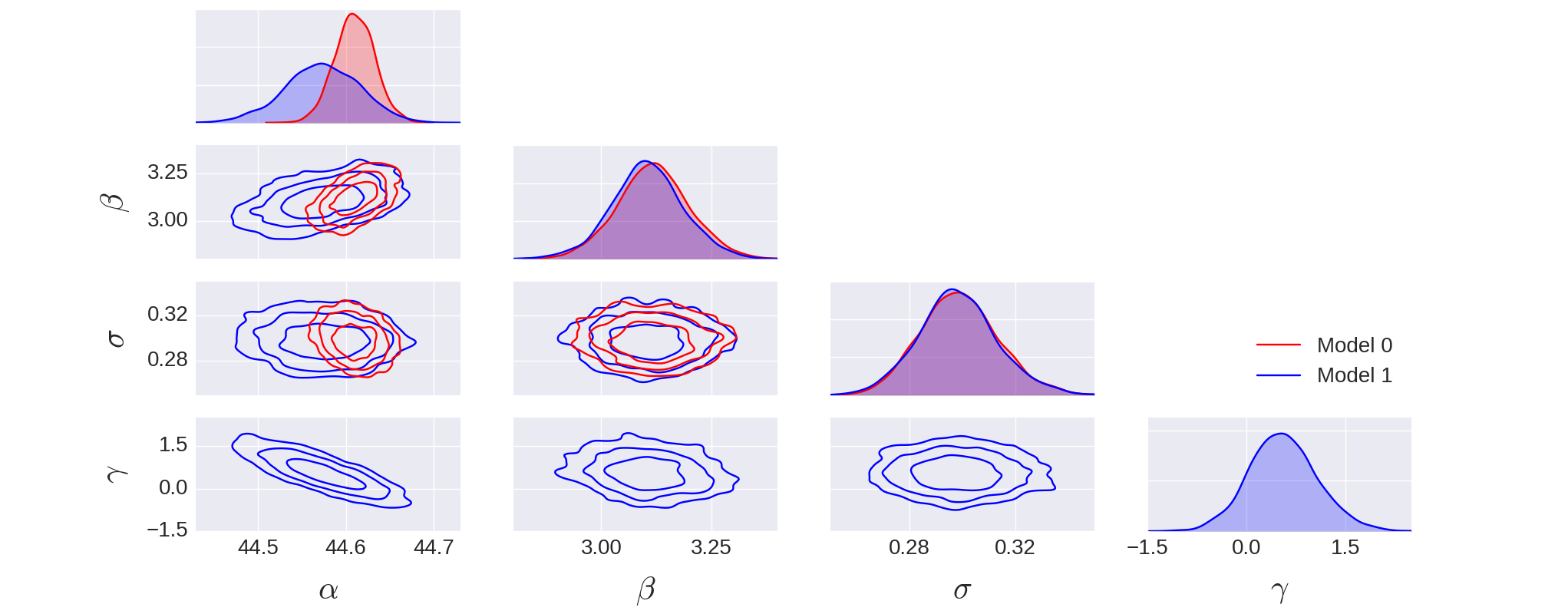}\\
\includegraphics[width=17.2cm]{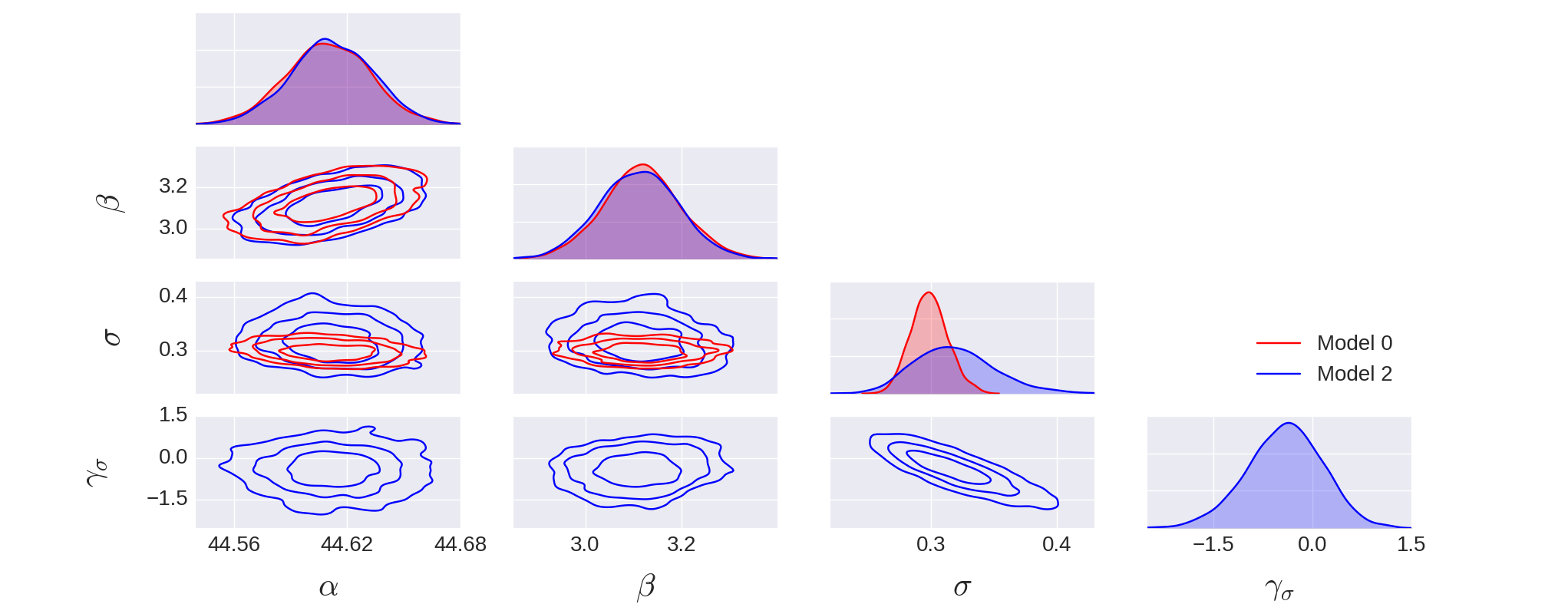}\\
\includegraphics[width=17.2cm]{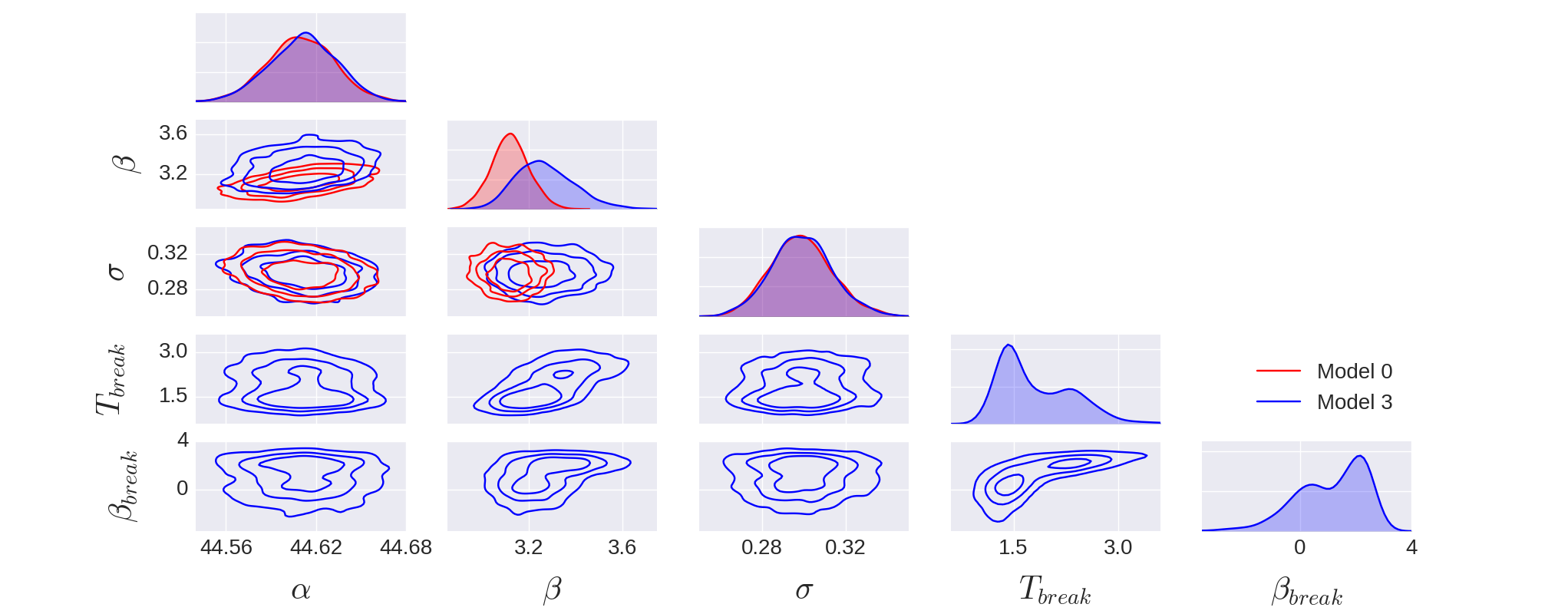}
\caption{Comparison of the one- and two-dimensional marginalised posterior distributions for models 1, 2 and 3, with those for model 0 (in red), respectively in the upper, middle and lower panels. Note that the horizontal axis that is associated with each one-dimensional marginalised posterior distribution is bellow it, at the bottom of each plot. The two-dimensional contours enclose 0.50 (inner), 0.80 (middle) and 0.95 (outer) of the total probability.}
\label{fig_model_0123}
\end{figure*}

We will now see what happens when model complexity is increased, by adding to the simplest model one of the following sets of hyper-parameters: $\gamma$, which allows the normalisation to evolve with time (model 1); $\gamma_\sigma$, which permits time-evolution of the scatter (model 2); $T_{\rm break}$ and $\beta_{\rm break}$, respectively, the transition temperature and the value of $\beta$ below $T_{\rm break}$ (model 3), associated with a possible change-point. The resulting regression lines can be seen in Figs. \ref{fig_model_1} (upper panel), \ref{fig_model_2} and \ref{fig_model_3}, respectively. In Fig. \ref{fig_model_1} we also show (lower panel) the results obtained when no sample selection effects are considered, which we label as model version 1u. In Fig. \ref{fig_model_0123} we compare the one- and two-dimensional marginalised posterior distributions for models 1, 2 and 3, with those for model 0.

\begin{figure*}
\includegraphics[width=\columnwidth]{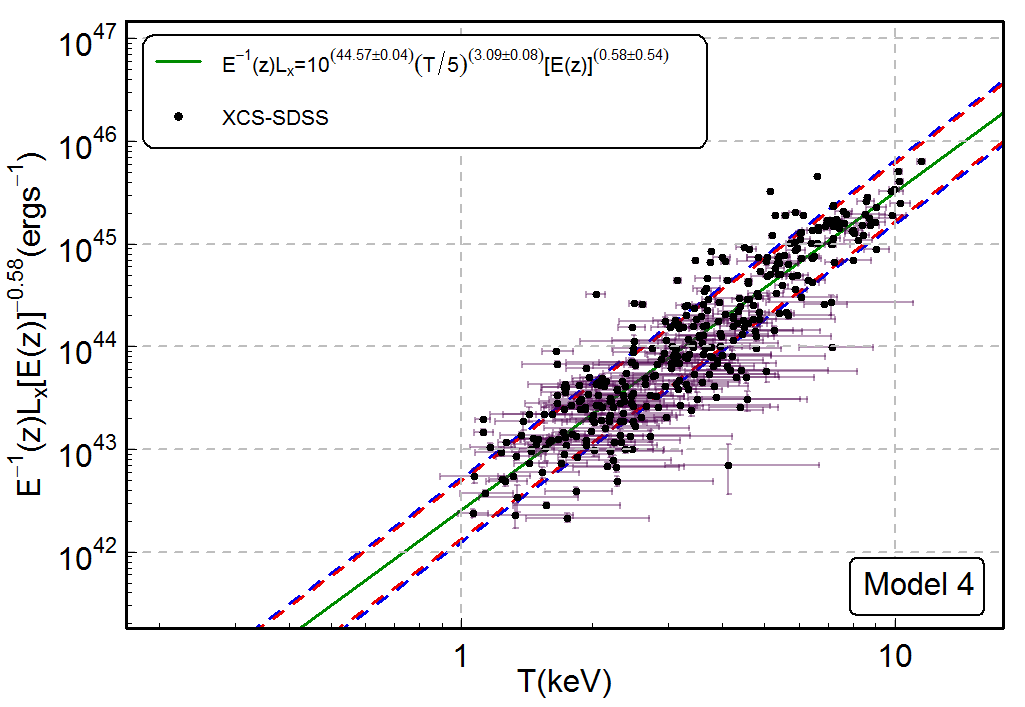}
\includegraphics[width=\columnwidth]{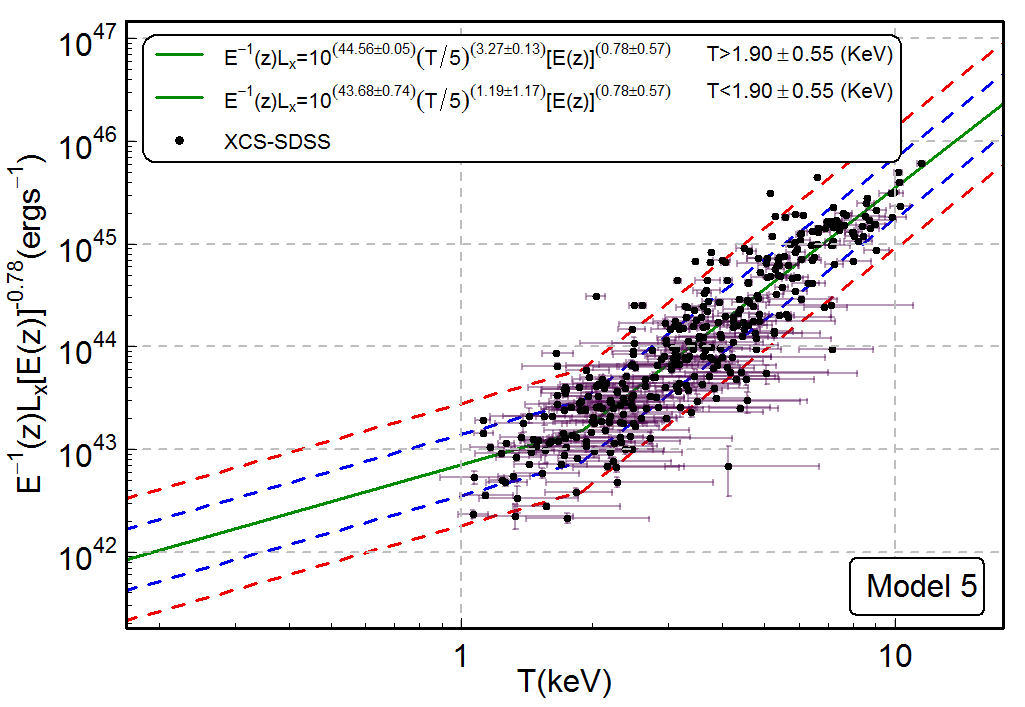}
\includegraphics[width=\columnwidth]{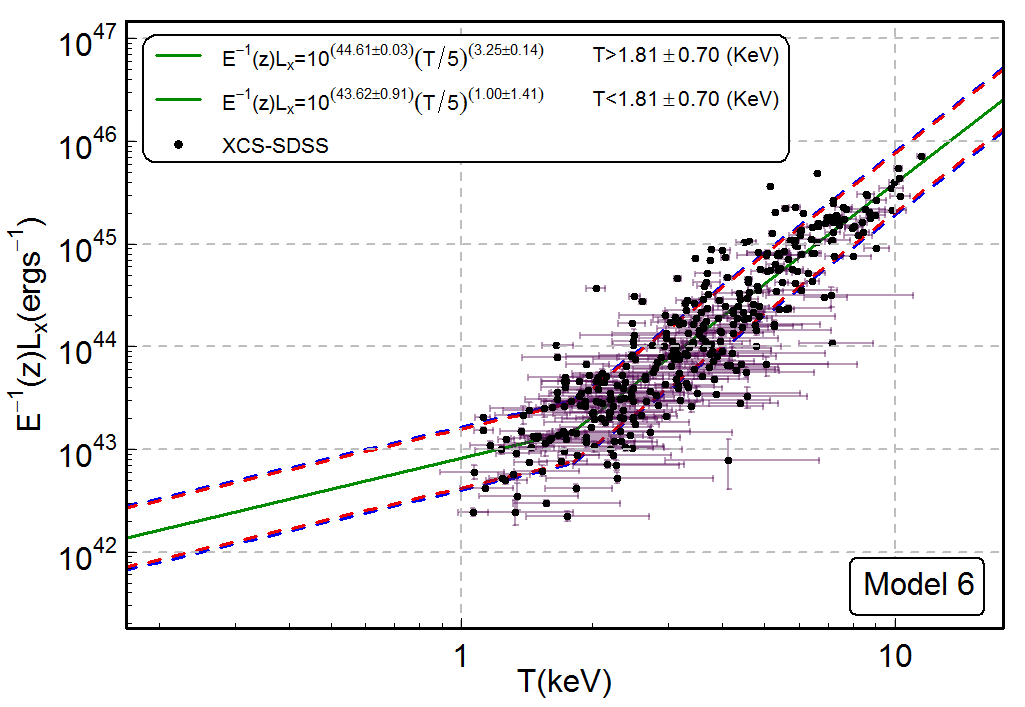}
\includegraphics[width=\columnwidth]{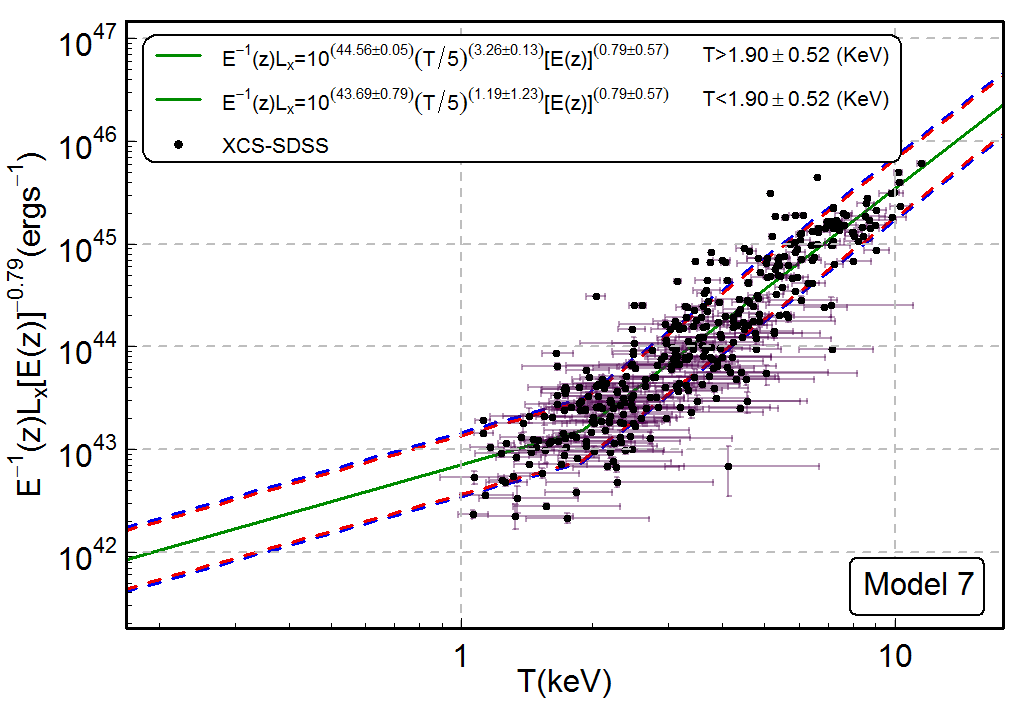}
\caption{The \lxt relation assuming models 4 (top, left), 5 (top, right), 6 (bottom, left) and 7 (bottom, right). The prior distribution for the temperature is always modelled through a time-evolving Gaussian. In the case of model 5, the amplitude of the intrinsic vertical scatter about the regression line is indicated by the dashed lines ($1\sigma$, inner blue; $2\sigma$, outer red). For models 4, 6 and 7, dashed lines indicate the amplitude of the intrinsic vertical scatter about the regression line only at the $1\sigma$ level, for the minimum (0.1, outer blue) and maximum (0.6, inner red) sample redshifts.}
\label{fig_model_4567}
\end{figure*}

\begin{figure*}
\includegraphics[width=17.4cm]{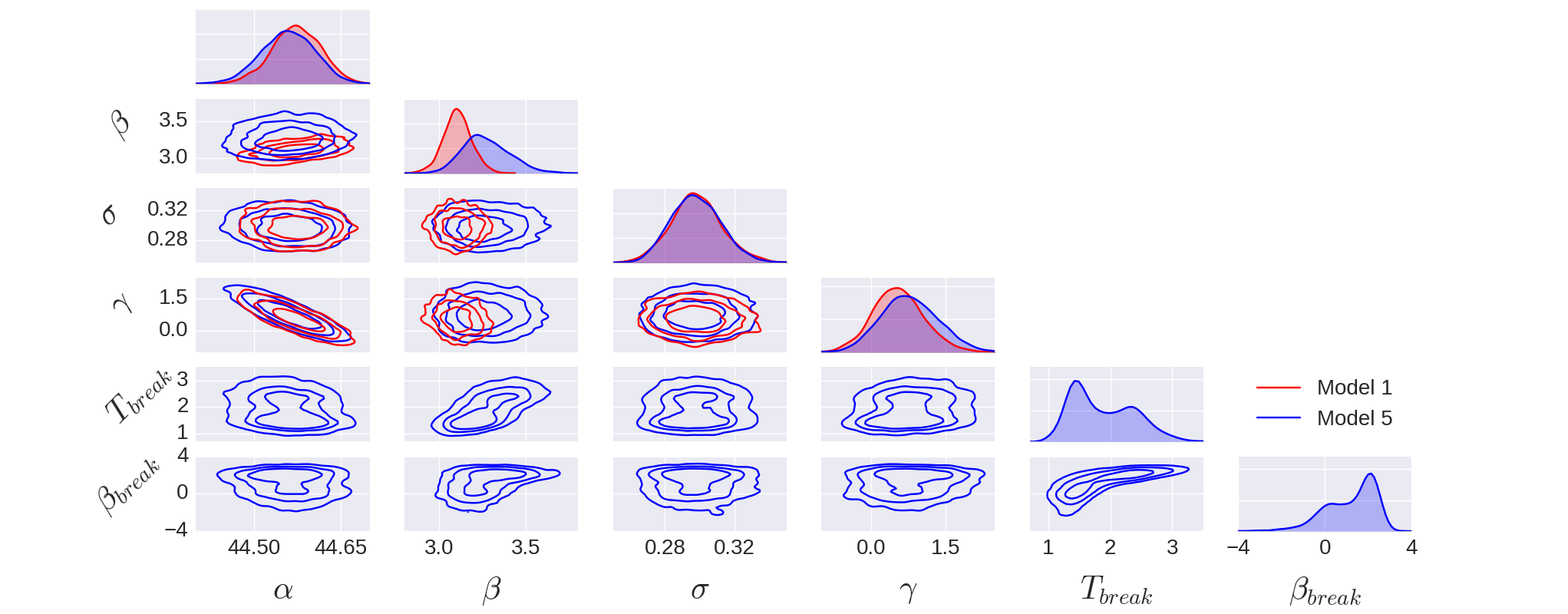}
\includegraphics[width=17.4cm]{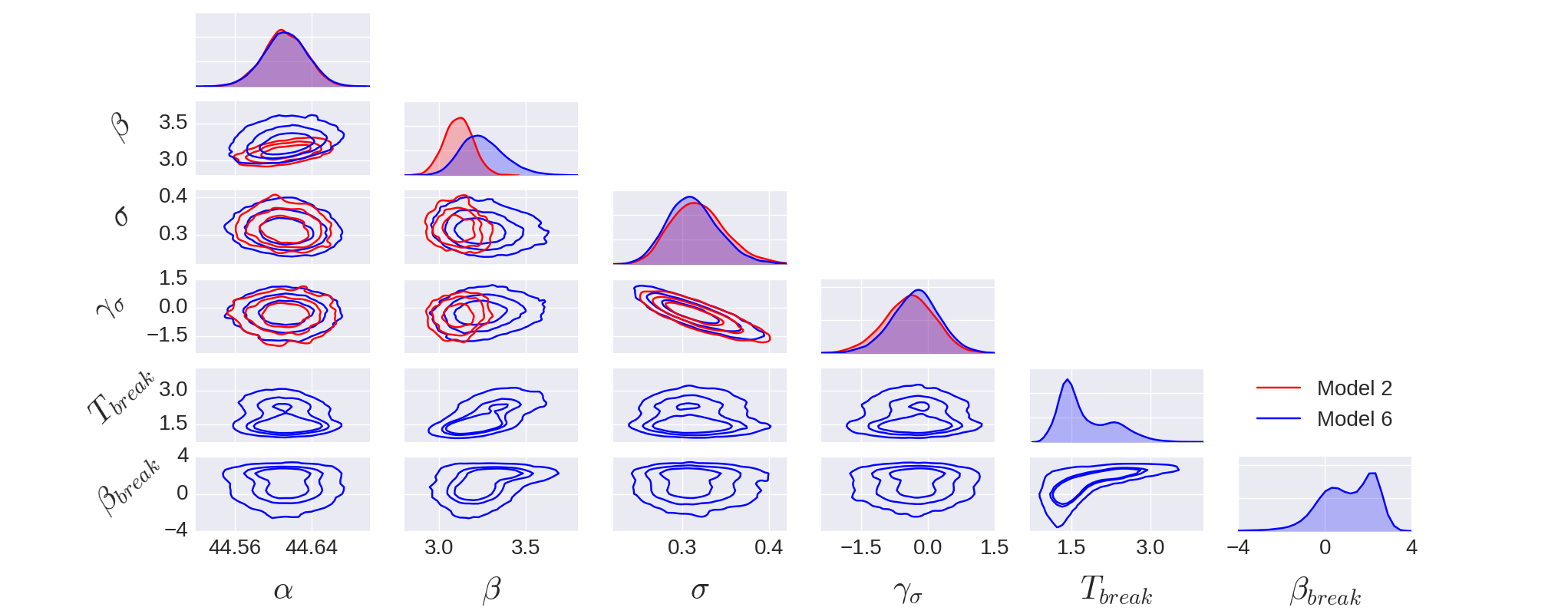}
\includegraphics[width=17.4cm]{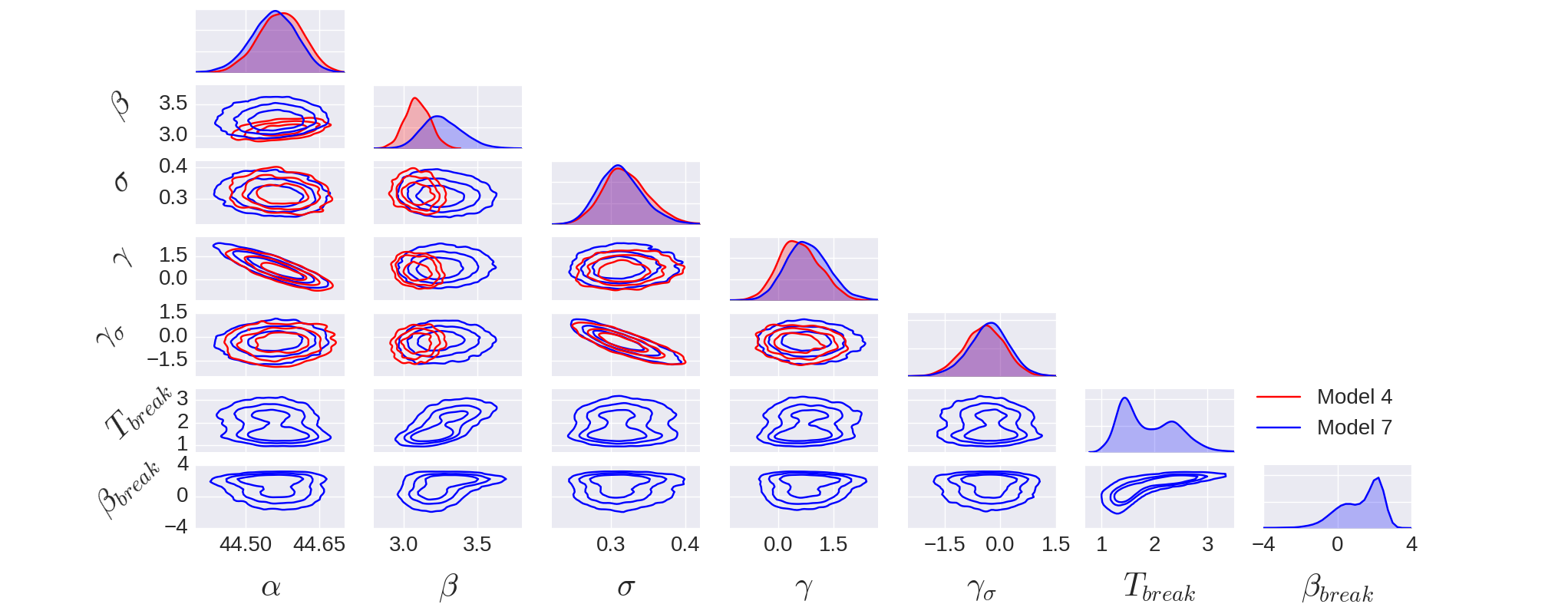}
\caption{Comparison of the one- and two-dimensional marginalised posterior distributions for models 1 vs. 5 (upper panel), 2 vs. 6 (middle panel), and 4 vs. 7 (lower panel). The two-dimensional contours enclose 0.50 (inner), 0.80 (middle) and 0.95 (outer) of the total probability.}
\label{fig_model_124567}
\end{figure*}

The even more complex models 4, 5, 6 and 7, include all possible combinations of the hyper-parameters introduced in the preceding models. The respective regression lines are represented in Fig. \ref{fig_model_4567}. The one- and two-dimensional marginalised posterior distributions for models 1 vs. 5, 2 vs. 6, and 4 vs. 7, are compared in Fig. \ref{fig_model_124567}, thus highlighting the impact of introducing a change-point in the \lxt relation. 

The posterior distributions for the hyper-parameters $T_{\rm break}$ and $\beta_{\rm break}$ rapidly decrease towards zero when approaching $1$ keV and $4$, respectively. This happens for all models, i.e. 3, 5 , 6, and 7, which include these hyper-parameters, as can be seen in Figs. \ref{fig_model_0123} and \ref{fig_model_124567}. The lower limit of $1$ keV on $T_{\rm break}$ is driven by the prior distribution assumed for this hyper-parameter, which imposed a prior probability of zero for $T_{\rm break}$ smaller than $1$ keV. It would not make sense to assume otherwise, because all objects in our sample have observed temperatures greater than $1$ keV. Only the addition to the sample of objects with temperatures below $1$ keV would allow for a better understanding of the behaviour of the posterior distribution for $T_{\rm break}$ close to and below $1$ keV. Nevertheless, given that its mode is close to $1.5$ keV, adding such objects to the sample should not change significantly the location of the highest posterior probability region associated with $T_{\rm break}$, and hence its expected value. On the other hand, the upper limit of $4$ on $\beta_{\rm break}$ is mostly driven by the data. Given that the normalization of the $\log L_{\rm X}$ - $\log T$ relation is strongly constrained, values for $\beta_{\rm break}$ approaching $4$ incur a rapidly growing penalty due to a regression line that lies increasingly far away from objects with low $T$ and relatively high $L_{\rm X}$. These have $T$ values with much smaller measurement uncertainties than objects with low $T$ and relatively low $L_{\rm X}$, thus regression lines that lie equally far away from the later, i.e. characterised by $\beta_{\rm break}$ values approaching $0$, are not as penalized, leading to a slower decrease in the posterior probability for $\beta_{\rm break}$ left of its mode.

In Table \ref{table1}, we provide the estimated median value and symmetric (or equal-tailed) credible intervals, centred on the median, for the marginal posterior distributions of each hyper-parameter of interest associated with the models being considered. In Table \ref{table2}, we provide the expected value (i.e. mean), standard deviation and skewness for the same marginal posterior distributions. As can be seen, some of the marginal posterior distributions deviate significantly from a Gaussian distribution, for which the skewness is zero. For these distributions the median and mean do not coincide, which is why we decided to include more information about the marginal posterior distributions than usual. The first three (raw) cumulants of a given probability distribution coincide, respectively, with the mean, the variance (i.e. the square of the standard deviation), and the product of the skewness with the cube of the standard deviation. The values of these cumulants can be used to approximate each marginal posterior distribution through an Edgeworth expansion around a Gaussian distribution with the same mean and variance, for example using the R package PDQutils available both at CRAN\footnotemark\footnotetext{\url{https://cran.r-project.org/web/packages/PDQutils/}} and GitHub\footnotemark.\footnotetext{\url{https://github.com/shabbychef/PDQutils/}}

\begin{table*}
\caption{The median and symmetric credible intervals, centred on the median, that contain 68.3\% and 95.4\% of the probability for the marginal posterior distributions of the interesting hyper-parameters associated with each model. After each median value, are represented the numbers that need to be added and subtracted from it, in order to obtain the limits associated with the respective credible interval.}
\label{table1}
\def\arraystretch{1.5}
\begin{tabular}{|c|c|c|c|c|c|c|c|}
\hline
Model & $\alpha$ & $\beta$ & $\sigma$ & $\gamma$ & $\gamma_\sigma$ & $T_{\rm break}$ & $\beta_{\rm break}$\\
\hline
$0{\rm u}$ & $44.61\,_{-0.02\,-0.04}^{+0.02\,+0.04}$ & $3.00\,_{-0.07\,-0.14}^{+0.07\,+0.15}$ & $0.30\,_{-0.01\,-0.03}^{+0.01\,+0.03}$ & $0$ & $0$ & $-$ & $-$\\
$0$ & $44.61\,_{-0.02\,-0.05}^{+0.02\,+0.05}$ & $3.12\,_{-0.08\,-0.16}^{+0.08\,+0.16}$ & $0.30\,_{-0.01\,-0.03}^{+0.02\,+0.03}$ & $0$ & $0$ & $-$ & $-$\\
$1{\rm u}$ & $44.56\,_{-0.04\,-0.09}^{+0.04\,+0.08}$ & $2.99\,_{-0.08\,-0.16}^{+0.08\,+0.15}$ & $0.30\,_{-0.02\,-0.03}^{+0.01\,+0.03}$ & $0.73\,_{-0.53\,-0.96}^{+0.56\,+1.16}$ & $0$ & $-$ & $-$\\
$1$ & $44.57\,_{-0.04\,-0.09}^{+0.04\,+0.08}$ & $3.10\,_{-0.08\,-0.17}^{+0.08\,+0.17}$ & $0.30\,_{-0.01\,-0.03}^{+0.01\,+0.03}$ & $0.55\,_{-0.49\,-1.01}^{+0.52\,+1.09}$ & $0$ & $-$ & $-$\\
$2$ & $44.61\,_{-0.02\,-0.04}^{+0.02\,+0.04}$ & $3.11\,_{-0.08\,-0.17}^{+0.08\,+0.16}$ & $0.32\,_{-0.03\,-0.05}^{+0.03\,+0.07}$ & $0$ & $-0.41\,_{-0.58\,-1.21}^{+0.55\,+1.07}$ & $-$ & $-$\\
$3$ & $44.61\,_{-0.02\,-0.05}^{+0.02\,+0.04}$ & $3.26\,_{-0.11\,-0.21}^{+0.14\,+0.29}$ & $0.30\,_{-0.01\,-0.03}^{+0.01\,+0.03}$ & $0$ & $0$ & $1.65\,_{-0.31\,-0.52}^{+0.77\,+1.34}$ & $1.40\,_{-1.56\,-3.12}^{+0.95\,+1.40}$\\
$4$ & $44.57\,_{-0.05\,-0.09}^{+0.04\,+0.08}$ & $3.09\,_{-0.08\,-0.16}^{+0.09\,+0.17}$ & $0.32\,_{-0.03\,-0.05}^{+0.03\,+0.07}$ & $0.55\,_{-0.49\,-0.96}^{+0.61\,+1.14}$ & $-0.42\,_{-0.56\,-1.15}^{+0.53\,+1.06}$ & $-$ & $-$\\
$5$ & $44.56\,_{-0.05\,-0.09}^{+0.05\,+0.09}$ & $3.26\,_{-0.12\,-0.22}^{+0.15\,+0.31}$ & $0.30\,_{-0.01\,-0.03}^{+0.02\,+0.03}$ & $0.76\,_{-0.55\,-1.06}^{+0.61\,+1.20}$ & $0$ &$1.79\,_{-0.42\,-0.64}^{+0.66\,+1.18}$ & $1.50\,_{-1.53\,-3.02}^{+0.79\,+1.24}$\\
$6$ & $44.61\,_{-0.02\,-0.05}^{+0.02\,+0.04}$ & $3.24\,_{-0.11\,-0.21}^{+0.14\,+0.31}$ & $0.31\,_{-0.03\,-0.05}^{+0.03\,+0.07}$ & $0$ & $-0.27\,_{-0.55\,-1.18}^{+0.52\,+1.06}$ & $1.58\,_{-0.27\,-0.50}^{+0.80\,+1.48}$ & $1.15\,_{-1.37\,-3.00}^{+1.17\,+1.68}$\\
$7$ & $44.57\,_{-0.04\,-0.09}^{+0.04\,+0.08}$ & $3.30\,_{-0.13\,-0.23}^{+0.14\,+0.29}$ & $0.32\,_{-0.03\,-0.05}^{+0.03\,+0.07}$ & $0.73\,_{-0.45\,-0.87}^{+0.48\,+0.97}$ & $-0.64\,_{-0.54\,-1.11}^{+0.50\,+0.98}$ & $2.29\,_{-0.82\,-1.08}^{+0.47\,+1.12}$ & $2.03\,_{-1.49\,-3.07}^{+0.44\,+0.78}$\\
\hline
\end{tabular}
\end{table*}

\begin{table*}
\caption{The mean, standard deviation and skewness for the marginal posterior distributions of the interesting hyper-parameters associated with each model.}
\label{table2}
\def\arraystretch{1.5}
\begin{tabular}{|c|c|c|c|c|c|c|c|}
\hline
Model & $\alpha$ & $\beta$ & $\sigma$ & $\gamma$ & $\gamma_\sigma$ & $T_{\rm break}$ & $\beta_{\rm break}$\\
\hline
$0{\rm u}$ & $44.61, 0.02, 0.03$ & $3.00, 0.08, 0.04$ & $0.30, 0.01, 0.17$ & $0$ & $0$ & $-$ & $-$\\
$0$ & $44.61, 0.02, -0.02$ & $3.12, 0.08, 0.06$ & $0.30, 0.01, 0.18$ & $0$ & $0$ & $-$ & $-$\\
$1{\rm u}$ & $44.56, 0.04, -0.14$ & $2.99, 0.08, -0.03$ & $0.30, 0.01, 0.15$ & $0.75, 0.54, 0.22$ & $0$ & $-$ & $-$\\
$1$ & $44.57, 0.04, -0.12$ & $3.11, 0.08, -0.01$ & $0.30, 0.02, 0.21$ & $0.56, 0.51, 0.20$ & $0$ & $-$ & $-$\\
$2$ & $44.61, 0.02, -0.04$ & $3.11, 0.08, 0.01$ & $0.32, 0.03, 0.51$ & $0$ & $-0.42, 0.57, -0.18$ & $-$ & $-$\\
$3$ & $44.61, 0.02, -0.13$ & $3.27, 0.12, 0.44$ & $0.30, 0.01, 0.13$ & $0$ & $0$ & $1.83, 0.56, 1.56$ & $1.10, 1.31, -1.44$\\
$4$ & $44.57, 0.04, -0.11$ & $3.09, 0.08, 0.03$ & $0.32, 0.03, 0.41$ & $0.58, 0.54, 0.17$ & $-0.43, 0.55, -0.06$ & $-$ & $-$\\
$5$ & $44.56, 0.05, -0.12$ & $3.27, 0.14, 0.41$ & $0.30, 0.01, 0.23$ & $0.79, 0.57, 0.16$ & $0$ &$1.90, 0.55, 1.64$ & $1.19, 1.17, -0.91$\\
$6$ & $44.61, 0.03, -5.36$ & $3.26, 0.14, 2.00$ & $0.31, 0.03, 0.45$ & $0$ & $-0.28, 0.55, -0.14$ & $1.81, 0.70, 4.57$ & $1.00, 1.41, -2.41$\\
$7$ & $44.56, 0.05, -0.20$ & $3.26, 0.13, 0.38$ & $0.31, 0.03, 0.47$ & $0.80, 0.57, 0.22$ & $-0.31, 0.55, -0.16$ & $1.90, 0.52, 0.55$ & $1.20, 1.23, -2.00$\\
\hline
\end{tabular}
\end{table*}


It is clear from both Tables \ref{table1} and \ref{table2}, that the introduction of a Gaussian (mixture) prior mainly affects the expected value for $\beta$. This can be understood by first noticing that multiplying the likelihood by a Gaussian effectively leads to a re-weighting of the sample data, where the data points with values for $\log(T/5)$ closer to the Gaussian expected value have their relative importance down-weighted with respect to the others. Further, for all models considered, the Gaussian prior expected value is always around $-0.26$ (2.75 keV). Although that is close to the mode ($-0.325$, 2.37 keV) of the sample distribution for $\log(T/5)$, this has a significant positive skew, as can be seen in Fig. \ref{zLT_histograms}, with median and mean equal to $-0.17$ (3.38 keV). Therefore, the introduction of such a Gaussian (mixture) prior effectively leads to a re-weighting of the data points in favour of those located in the high temperature tail of the sample distribution. Because these prefer a higher value for $\beta$, as can be seen by what happens to the marginal posterior distribution for this parameter in the models where a change-point is introduced in the $L_{\rm X}-T$ relation, the re-weighting in their favour leads to the observed increase in the expected value for $\beta$ when a Gaussian (mixture) prior is introduced (in model 0 vs 0u, and in model 1 vs 1u).

The consequence of completely removing the data points associated with lower temperatures from the estimation of $\beta$ can be seen when a change-point is introduced, in model 3 vs 0, in model 5 vs 1, model 6 vs 2 and in model 7 vs 4. In every case the expected value for $\beta$ increases, by a similar margin as a result of introducing a Gaussian (mixture) prior. That happens because, as previously mentioned, the data points associated with lower temperatures prefer a smaller value for $\beta$, as can be seen from the expected values for $\beta_{\rm break}$.

Finaly, the introduction of the time-evolution parameters, $\gamma$ and $\gamma_\sigma$, affects most importantly the expected values for the base parameters to which they are associated, respectively $\alpha$ and $\sigma$, as can be seen by comparing the expected values for the models 1u vs 0u, 1 vs 0, 4 vs 2, 5 vs 3 and 7 vs 6 (with respect to $\gamma$), and models 2 vs 0, 4 vs 1, 6 vs 3 and 7 vs 5 (with respect to $\gamma_\sigma$), and taking into account the relative uncertainty associated with those parameters. Although not as strongly, the values for the change-point associated parameters, $T_{\rm break}$ and $\beta_{\rm break}$, especially the former, are also affected by the introduction of $\gamma$ and/or $\gamma_\sigma$. 

In Table \ref{table3}, we provide the estimates for WBIC and WAIC, as well as the associated sample standard deviations, with respect to each model considered. The results clearly indicate strong support for the introduction of a Gaussian (mixture) prior to model the joint effect of the population-level temperature function and any sample selection effects not explicitly taken into consideration (like those associated with SDSS-DR8-redMaPPer catalogue assembly, in our case). However, the data does not seem to hold enough information to differentiate between all favoured models, 0 to 7, in that respect. All these models have associated WBIC and WAIC estimates which overlap well within the respective sample standard deviations. This suggests that another, statistically equivalent, sample taken from the same underlying population could easily lead to a different ranking of the models 0 to 7, based on their associated WBIC and WAIC estimates for such sample. Interestingly, model 4 is that which gathers more support from the data, according to both the WBIC and WAIC model comparison methods. But, these two methods differ with respect to which individual feature, among those that make models 1, 2 and 3 distinct from model 0, is better supported by the data: WBIC suggests it is time-evolution of the intrinsic scatter about the $L_{\rm X}-T$ relation, parametrised by $\gamma_\sigma$; while WAIC prefers time-evolution of the normalization of the $L_{\rm X}-T$ relation beyond the self-similar expectation, parametrised by $\gamma$. However, both methods agree that the individual feature with least support in the data is an abrupt transition or change-point in the $L_{\rm X}-T$ relation, parametrised through $T_{\rm break}$ and $\beta_{\rm break}$. Assuming {\it a priori} any of models 0 to 7 to be true, the data consistently suggests a positive deviation with respect to self-similar evolution for the $L_{\rm X}-T$ relation (i.e. at fixed temperature the X-ray luminosity increases with redshift), a negative redshift evolution in the intrinsic scatter about the $L_{\rm X}-T$ relation (i.e. the scatter decreases with redshift), as well as a change-point location just below 2 keV and an increase in the exponent for the $L_{\rm X}-T$ relation when moving from the group to the cluster regime.

\begin{table}
\caption{Estimates for WBIC, its sample standard deviation, $\sigma_{ \rm WBIC}$, WAIC, and its sample standard deviation, $\sigma_{\rm WAIC}$, with respect to each model.}
\label{table3}
\begin{center}
\begin{tabular}{|c|c|c|c|c|}
\hline
Model & WBIC & $\sigma_{ \rm WBIC}$ & WAIC & $\sigma_{ \rm WAIC}$\\
\hline
$0{\rm u}$ & $2052$ & $38$ & $6.51$ & $0.11$\\
$0$ & $-1278$ & $37$ & $-2.88$ & $0.11$\\
$1{\rm u}$ & $2054$ & $38$ & $6.52$ & $0.11$\\
$1$ & $-1277$ & $37$ & $-2.88$ & $0.11$\\
$2$ & $-1277$ & $37$ & $-2.88$ & $0.11$\\
$3$ & $-1277$ & $37$ & $-2.85$ & $0.11$\\
$4$ & $-1282$ & $37$ & $-2.89$ & $0.11$\\
$5$ & $-1280$ & $37$ & $-2.85$ & $0.11$\\
$6$ & $-1278$ & $37$ & $-2.85$ & $0.11$\\
$7$ & $-1281$ & $37$ & $-2.85$ & $0.11$\\
\hline
\end{tabular}
\end{center}
\end{table}

\begin{figure}
\includegraphics[width=\columnwidth]{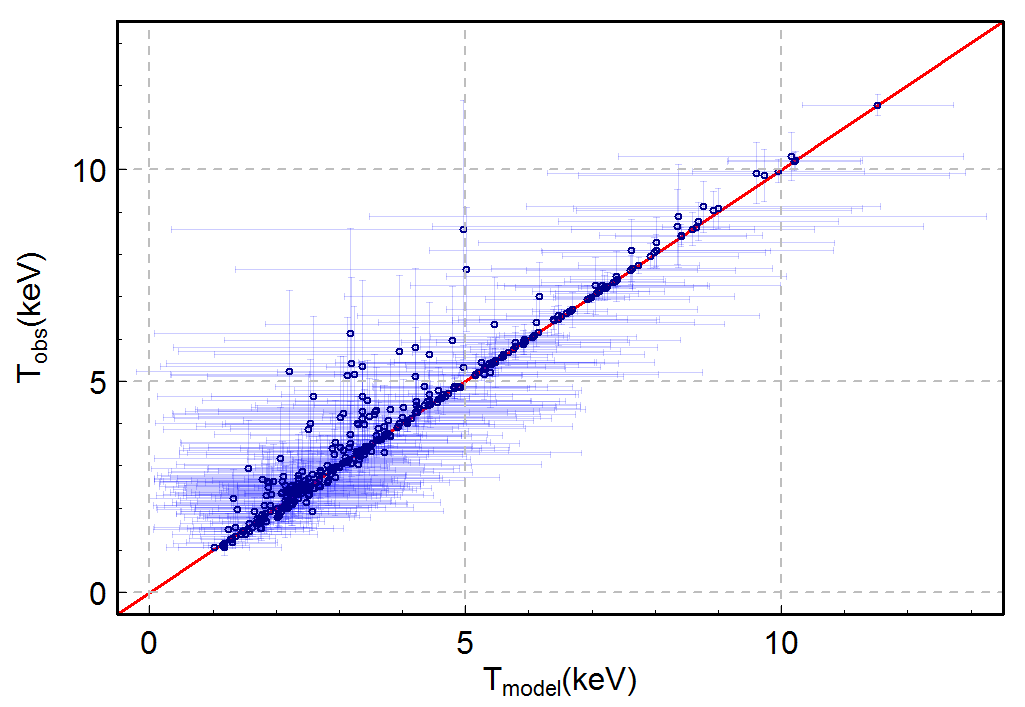}
\includegraphics[width=\columnwidth]{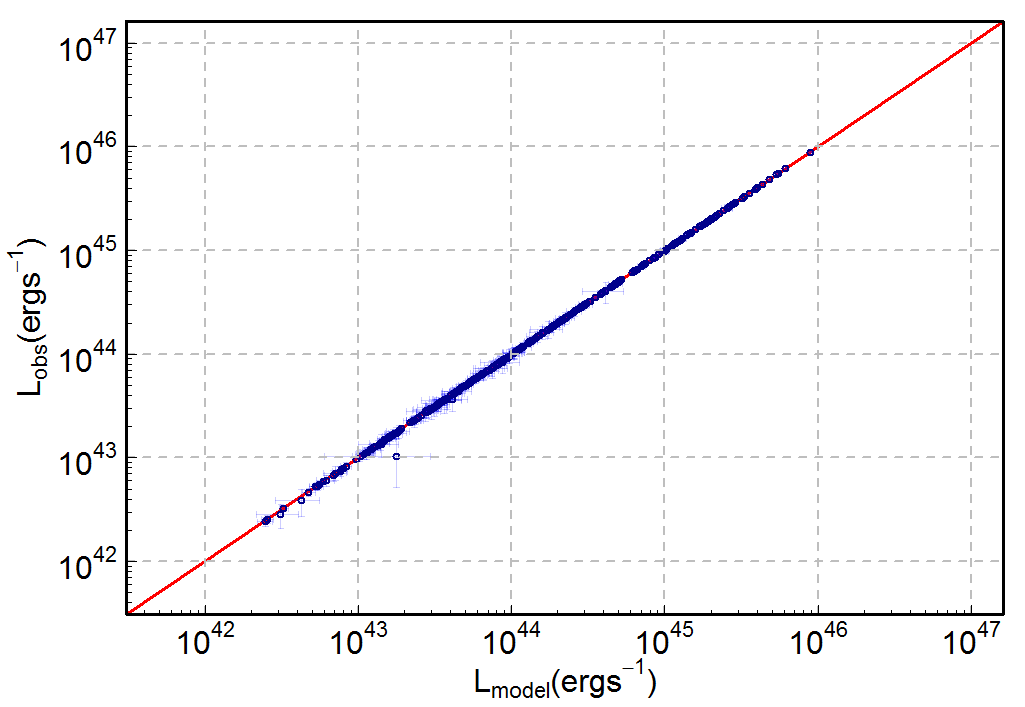}
\caption{Comparison, for all 353 systems in our sample, of the expected values and associated standard deviations associated with: the observed temperatures and X-ray luminosities, $T_{\rm obs}$ and $L_{\rm obs}$; the temperatures and X-ray luminosities, $T_{\rm model}$ and $L_{\rm model}$, obtained assuming {\it a priori} the $L_{\rm X}-T$ relation to be described by model 4.}
\label{lt_model_4}
\end{figure}

In Fig. \ref{lt_model_4}, we compare, for all 353 systems in our sample, the expected values and associated standard deviations associated with: the observed temperatures and X-ray luminosities, $T_{\rm obs}$ and $L_{\rm obs}$; the temperatures and X-ray luminosities, $T_{\rm model}$ and $L_{\rm model}$, obtained through LIRA, assuming {\it a priori} the $L_{\rm X}-T$ relation to be described by model 4. In the former case, expected values and standard deviations were estimated by approximating the probability distributions for the temperatures and X-ray luminosities that result from our X-ray analysis pipeline as univariate split Gaussian distributions \citep[][]{Villani_2006}, and using the modes and 68\% confidence intervals presented in Table \ref{table_data}. Further, the hierarchical modelling framework, implemented through LIRA, actually yields expected values and standard deviations for the (base 10) logarithms of $T_{\rm model}$ and $L_{\rm model}$. However, the posterior probability distributions for these quantities are nearly Gaussian, thus implying that the posterior probability distributions for $T_{\rm model}$ and $L_{\rm model}$ can be well approximated by (base 10) log-normal distributions. This means that the expected values, $\mu\,_{\rm model}$, and associated standard deviations, $\sigma\,_{\rm model}$, for $T_{\rm model}$ and $L_{\rm model}$ can be estimated from the respective quantities, $\mu\,_{\rm model,\,\log}$ and $\sigma\,_{\rm model,\,\log}$, for $\log T_{\rm model}$ and $\log L_{\rm model}$, through
\begin{equation}
\mu\,_{\rm model}=10^{\;(\mu\,_{\rm model,\,\log}+\frac{\ln10}{2}\sigma\,_{\rm model, \log}^2)}\,,
\label{eq18}
\end{equation}

\noindent and
\begin{equation}
\sigma\,_{\rm model}=\mu\,_{\rm model}\sqrt{10^{\;[\ln(10)\,\sigma\,_{\rm model,\,\log}^2]}-1}\,,
\label{eq19}
\end{equation}

\noindent where, as usual, $\ln$ stands for the natural logarithm. 

\begin{figure}
\includegraphics[width=\columnwidth]{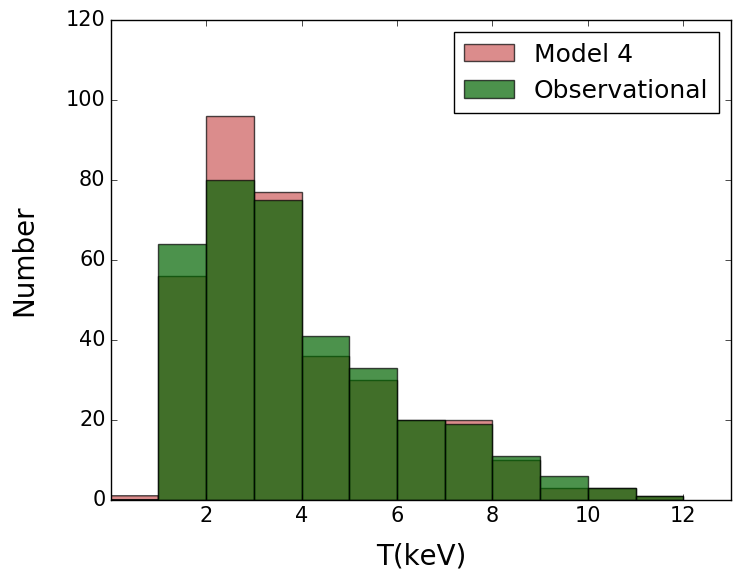}
\includegraphics[width=\columnwidth]{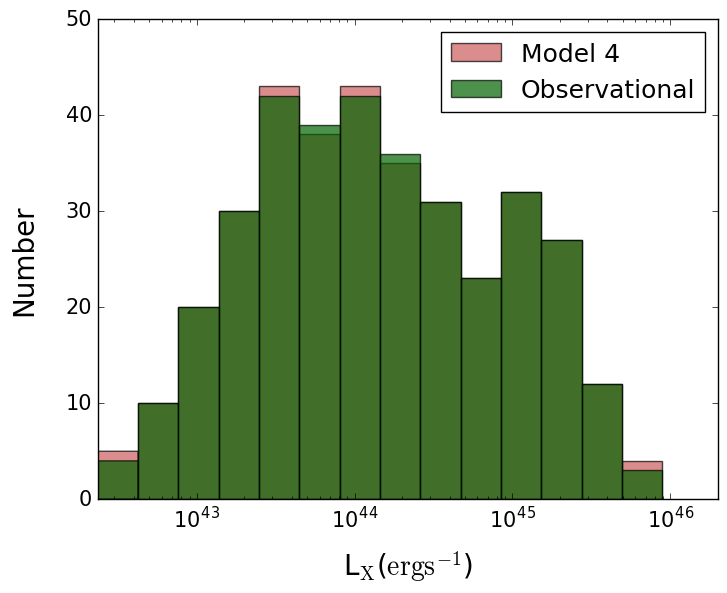}
\caption{Comparison of the sample distributions of most probable values for the observed temperatures and X-ray luminosities, with those obtained through LIRA, assuming {\it a priori} the $L_{\rm X}-T$ relation to be described by model 4.}
\label{lt_hist_model_4}
\end{figure}

In Fig. \ref{lt_hist_model_4}, we compare the sample distributions of most probable values for the observed temperatures and X-ray luminosities, with those obtained through LIRA, assuming {\it a priori} the $L_{\rm X}-T$ relation to be described by model 4. In the later case, the most probable values (i.e. modes) are given by

\begin{equation}
Mo\,_{\rm model}=10^{\;(\mu\,_{\rm model,\,\log}-\ln10\;\sigma\,_{\rm model, \log}^2)}\,,
\label{eq20}
\end{equation}

\noindent and are listed in a supplementary table available with the online edition of this article, together with the associated expected values and standard deviations, with respect to the temperature and X-ray luminosity posterior distributions for all systems in our sample. The first 10 rows are shown in Table \ref{table_model4}.

\begin{table*}
\caption{The first 10 rows of a supplementary table available with the online edition of this article, providing the XCS ID, most probable values for the temperature, $Mo\,_{T,\,{\rm model}}$ (in units of keV), and X-ray bolometric luminosity within $R_{500}$, $Mo\,_{L_X,\,{\rm model}}$ (in units of $10^{44}$ erg s$^{-1}$), as well as the associated expected values, $\mu\,_{T,\,{\rm model}}$ and $\mu\,_{L_X,\,{\rm model}}$, and standard deviations, $\sigma\,_{T,\,{\rm model}}$ and $\sigma\,_{L_X,\,{\rm model}}$, obtained through LIRA, assuming {\it a priori} the $L_{\rm X}-T$ relation to be described by model 4, for all 353 sample systems.}
\label{table_model4}
\def\arraystretch{1.5}
\begin{tabular}{|c|c|c|c|c|c|c|}
\hline
XCS ID & $Mo\,_{T,\,{\rm model}}$ & $\mu\,_{T,\,{\rm model}}$ & $\sigma\,_{T,\,{\rm model}}$ & $Mo\,_{L,\,{\rm model}}$ & $\mu\,_{L,\,{\rm model}}$ & $\sigma\,_{L,\,{\rm model}}$\\
\hline
${\rm XMMXCS\;J140101.9+025238.3}$ & $6.64$ & $6.64$ & $0.13$ & $55.39$ & $55.39$ & $0.10$\\
${\rm XMMXCS\;J142601.0+374937.0}$ & $8.65$ & $8.65$ & $0.53$ & $31.88$ & $31.88$ & $0.22$\\ 
${\rm XMMXCS\;J172009.8+263725.8}$ & $6.15$ & $6.15$ & $0.21$ & $21.50$ & $21.50$ & $0.09$\\   
${\rm XMMXCS\;J224321.4-093550.2}$ & $7.24$ & $7.24$ & $0.40$ & $33.11$ & $33.11$ & $0.11$\\
${\rm XMMXCS\;J212939.7+000516.9}$ & $5.29$ & $5.30$ & $0.27$ & $22.75$ & $22.75$ & $0.13$\\ 
${\rm XMMXCS\;J075124.2+173057.7}$ & $3.29$ & $3.29$ & $0.26$ & $2.86$ & $2.86$ & $0.02$\\  
${\rm XMMXCS\;J172227.0+320758.0}$ & $7.21$ & $7.21$ & $0.43$ & $27.57$ & $27.57$ & $0.19$\\
${\rm XMMXCS\;J104044.2+395711.1}$ & $3.74$ & $3.74$ & $0.18$ & $7.39$ & $7.39$ & $0.04$\\ 
${\rm XMMXCS\;J111253.4+132640.2}$ & $4.89$ & $4.89$ & $0.29$ & $6.12$ & $6.12$ & $0.04$\\
${\rm XMMXCS\;J024803.3-033143.4}$  & $3.77$ & $3.77$ & $0.19$ & $9.64$ & $9.64$ & $0.06$\\  
\hline
\end{tabular}
\end{table*}

As expected, assuming {\it a priori} a specific model for the $L_{\rm X}-T$ relation induces regression towards the mean, leading to a decrease in the variance of the temperature and X-ray luminosity sample distributions. This effect is significantly stronger for the former with respect to the later, as a result of the higher uncertainty level associated with the data-only based estimates for the temperatures. Substituting model 4 by any other model between 0 and 7 yields very similar results. On an individual system basis, the higher the overall uncertainty in data-only based estimates, the larger can be the difference between these and estimates that assume {\it a priori} a specific model for the $L_{\rm X}-T$ relation. This is well illustrated in the case of cluster XMMXCS J222855.3+203920.1, which stands out as an outlier in the $L_{\rm obs}$ versus $L_{\rm model}$ plot in fig. \ref{lt_model_4}. The significant difference between these two estimates stems from the large uncertainty in both the temperature and X-ray luminosity data-only based estimates for this system, together with a well below-average value for the later given the former, with respect to the sample distribution of such estimates.

\section{Discussion}
\label{s_discussion}

We have shown that the data considered, generally, does not support the existence of deviations from the standard modelling of the $L_{\rm X}-T$ cluster and group scaling relation, as a single power-law, with a time-evolving normalization that follows the self-similar expectation and a non-evolving scatter. In particular, among the extra features we considered could be added to this standard model, the existence of a change-point in the $L_{\rm X}-T$ relation, marking the transition between what would be the group and cluster regimes, is that which seems less justified given the data.

It has been shown by \citet[][]{Maughan_2014} that the exponent of the $L_{\rm X}-T$ scaling relation, $\beta$, can be written as the ratio of the exponents of the $L_{\rm X}-M$ and $T-M$ scaling relations, where $M$ stands for mass, respectively $\beta_{LM}$ and $\beta_{TM}$. These can be thought as more fundamental, given that the presence of mass is the most determining factor in the assembly of both groups and clusters, as the result of gravitational instability. In fact, under the self-similar assumption \citep[][]{Kaiser_1986}, groups and clusters are assumed to be structures that virialize under the influence of gravity alone. This leads to a self-similar expectation of $\beta_{TM}=2/3$. The derivation of the self-similar expectation for $\beta_{LM}$ is more complicated, and (usually) implies assuming that the bolometric luminosity is the result of bremsstrahlung emission only, as well as that both the total amount of gas and the way it is distributed do not depend on the mass of the structures being considered. These assumptions then lead to the self-similar expectation $\beta_{LM}=4/3$, and hence $\beta=2$. However, the contribution of line emission to the bolometric luminosity in structures with temperatures below about 2 keV cannot be neglected, and thus the self-similar expectation for both $\beta_{LM}$ and $\beta$ should be different in this temperature regime. Indeed, we have confirmed that the ratio between bremsstrahlung and total emission increases roughly with the square root of the temperature for all systems with temperatures below 2 keV in our sample, as proposed by \citet[][]{Zou_2016}. Hence, the total bolometric luminosity is in fact proportional to its bremsstrahlung component divided by the square root of the plasma temperature. The self-similar expectations for $\beta_{LM}$ and $\beta$ then become, respectively, $1$ and $3/2$, for structures with temperatures below about 2 keV \citep[see also][]{Sun_2012}. The self-similar expectation is thus of a steepening of the $L_{\rm X}-T$ relation when moving from the group to the cluster regime. 

However, most analysis suggest the values for $\beta_{LM}$, $\beta_{TM}$ and $\beta$ predicted under the self-similar assumption are disfavoured given the available cluster data \citep[e.g. see review by][]{Giodini_2013}. This is most probably the result of radiative cooling and AGN feedback acting alongside gravity during the assembly and subsequent evolution of groups and clusters, as shown by simulations \citep[e.g.][]{Short_2013, Planelles_2014, Barnes_2017, LeBrun_2017,Truong_2018}. Both effects seem to lead to an increase in both $\beta_{LM}$ and $\beta$, relatively to the self-similar expectation, with $\beta_{TM}$ decreasing proportionally less. If these effects were more important, with respect to those induced by gravity, in groups than in clusters, they would contribute to erasing some of the difference between the group and cluster self-similar expectations for $\beta_{LM}$ and $\beta$. Nevertheless, it would be somewhat surprising that, by coincidence, the stronger effects of radiative cooling and AGN feedback in groups versus clusters, would just compensate a completely unrelated effect due to the temperature dependence of line emission. Therefore, it is expectable {\it a priori} the existence of a change in the exponent of the $L_{\rm X}-T$ scaling relation, when moving from the group to the cluster regime, albeit possibly weaker than the self-similar expectation.

The fact that we do not find statistical support for the existence of a change in the exponent of the $L_{\rm X}-T$ scaling relation, could be due to its real absence or to a lack of sufficient information in the data to allow for its unambiguous identification. Nevertheless, if we indeed assume {\it a priori} a change-point in the $L_{\rm X}-T$ relation to exist, as in models 3, 5, 6 and 7, we do find concordance with the self-similar expectation of an increase in the expected value for the exponent of the $L_{\rm X}-T$ relation when moving from the group to the cluster regime. In the context of those models, the expected value for the $L_{\rm X}-T$ relation exponent in the group regime, around $1.1$, is even smaller than the self-similar expectation, although the difference is not statistically significant. On the other hand, the large positive difference between the expected value found here for the exponent in the cluster regime, close to $3.3$, and the self-similar expectation, is statistically robust and in agreement with most previous results \citep[e.g.][]{MantzScaling_2010, Mittal_2011, Maughan_2012, Takey_2013, Maughan_2014, Giles_2016, Mantz_2016}. It seems to be essentially the result of a steeper than self-similar $L_{\rm X}-M$ relation \citep[][]{Maughan_2014, Mantz_2016}, although there could be also some contribution from a shallower than self-similar $T-M$ relation \citep[][]{Maughan_2014, Lieu_2016, Mantz_2016}. A smaller value for the exponent in the group regime, with respect to the self-similar expectation, together with a higher value in the cluster regime, could indicate that either the $L_{\rm X}-M$ relation is shallower, or the $T-M$ relation is steeper, in the case of groups and again with respect to the self-similar expectation. For example, assuming the $T-M$ relation to be similar for both groups and clusters would imply that the exponent of the $L_{\rm X}-M$ relation for groups would be close to the ratio between the expected values for the exponents of the $L_{\rm X}-T$ relation for groups ($1.1$) and clusters ($3.3$) times the exponent of the $L_{\rm X}-M$ relation for clusters. Assuming a exponent around 1.6 for the cluster $L_{\rm X}-M$ relation \citep[][]{Maughan_2014}, which is 20\% higher than the self-similar expectation, would then imply the expected value for the exponent of the group $L_{\rm X}-M$ relation would be $0.5$, clearly shallower than the self-similar expectation, although compatible within the statistical uncertainties.

The expected values we obtain for the exponent of the $L_{\rm X}-T$ relation in the group regime, in the context of models 3, 5, 6 and 7, are significantly smaller than what was found by \citet[][]{Bharadwaj_2015}, who updated the analysis performed by \citet[][]{Eckmiller_2011}, and \citet[][]{Zou_2016}, but compatible within the statistical uncertainties. In Table \ref{table2} we compare the expected values found for the exponent of the $L_{\rm X}-T$ relation by \citet[][]{Bharadwaj_2015} and \citet[][]{Zou_2016} with those obtained here assuming models 3, 5, 6 and 7. We also present the expected values obtained when models 0u, 0, 1u, 1, 2 and 4 are applied to each of the sub-samples that result from splitting the objects in our full sample according to whether their observed (most probable) temperature is lower or higher than either 2 or 3 keV. In the case of the former there are 55 systems below the threshold, with 298 above, while for the later there are 142 systems below the threshold, and 211 above. We chose the 2 keV threshold because it is close to the expected value of $T_{\rm break}$ found for most of the models considered, but they cannot be directly compared. A hard selection threshold in observed temperature translates into a probabilistic boundary with respect to the model-inferred expected temperatures (and vice-versa), in the sense that some systems that end up in the e.g. lower observed temperature sub-sample actually have model-inferred expected temperatures higher than the value associated with the threshold (and vice-versa). The alternative 3 keV threshold was chosen because it leads to a observed temperature range, for the lower temperature sub-sample, similar to that associated with the group samples considered in \citet[][]{Bharadwaj_2015} and \citet[][]{Zou_2016}. In these analysis, a model for the $L_{\rm X}-T$ relation that is equivalent to our model 0u was assumed, and the expected values for the parameters of such model, $\theta_0$, were then determined, both with and without trying to take sample selection effects into account.

The expected values we obtain for the exponent of the $L_{\rm X}-T$ relation in the case of model 0u, are only $0.12$ ($T<2\;{\rm keV}$) and $0.35$ ($T<3\;{\rm keV}$), for the lower temperature sub-samples, raising to $2.73$ ($T>2\;{\rm keV}$) and $2.61$ ($T>3\;{\rm keV}$), for the higher temperature sub-samples. The former are significantly smaller than the exponents obtained by \citet[][]{Bharadwaj_2015} and \citet[][]{Zou_2016}, respectively $2.17$ and $2.80$, which suggests that the selection effects acting on the low temperature end of our sample are significantly different from those acting on their group samples. Such effects could be, for example, the result of a different ratio of strong cool core versus non-cool core \citep[][]{Bharadwaj_2015, Giles_2016, Zou_2016}, or relaxed vs. non-relaxed \citep[e.g.][]{Chon_2017}, systems among the samples. Allowing for deviations from self-similar redshfit evolution (model 1u) does not change significantly the expected values for the exponent of the $L_{\rm X}-T$ relation, for both temperature sub-samples. If we now attempt to correct for sample selection effects, by assuming a Gaussian prior distribution for our covariate, $\log(T/5)$, we then find for the exponent expected values, in the case of model 0, $2.55$ ($T<2\;{\rm keV}$) and $2.05$ ($T<3\;{\rm keV}$), and in the case of model 1, $1.84$ ($T<2\;{\rm keV}$) and $1.68$ ($T<3\;{\rm keV}$), for the lower temperature sub-samples. These values rise to $3.20$ ($T>2\;{\rm keV}$) and $3.06$ ($T>3\;{\rm keV}$), in the case of model 0, $3.20$ ($T>2\;{\rm keV}$) and $3.08$ ($T>3\;{\rm keV}$), in the case of model 1, for the higher temperature sub-samples. Therefore, as expected given the nature of the selection effects acting on group and cluster samples, talking these into account always leads to a significant increase in the expected value for the exponent of the $L_{\rm X}-T$ relation, as also found by \citet[][]{Bharadwaj_2015} and \citet[][]{Zou_2016}. This increase, as well as the larger associated uncertainty, is enough to bring our results for the $T<2\;{\rm keV}$ (but not for $T<3\;{\rm keV}$) sub-sample, in the context of model 0, to agreement with those obtained by \citet[][]{Bharadwaj_2015} and \citet[][]{Zou_2016}. The same conclusions are reached in the context of models which allow for possible redshift evolution beyond the self-similar expectation (models 1 and 4) and/or in the scatter (models 2 and 4), although the expected values for the exponent of the $L_{\rm X}-T$ relation become significantly smaller assuming models 1 and 4 than those inferred in the context of model 0. This difference seems to be the result of a preference, in the case of low temperature systems, for very strong positive redshift evolution beyond what is expected in the self-similar model.

\begin{table}
\caption{The exponent of the $L_{\rm X}-T$ relation estimated by \citet[][]{Bharadwaj_2015} and \citet[][]{Zou_2016} (usingY | X regression, {\em private communication}), compared with the expected values for the exponent we obtain assuming different models and samples.}
\label{table4}
\begin{tabular}{|l|c|}
\hline
Sample selection effects not considered& \\
\hline
\citet[][]{Bharadwaj_2015} & $2.17 \pm 0.26$ \\
\citet[][]{Zou_2016} & $2.80 \pm 0.22$ \\
{\rm Model} $0u$ ({\rm for} $T<2\;{\rm keV}$) & $0.12 \pm 0.10$ \\
{\rm Model} $0u$ ({\rm for} $T>2\;{\rm keV}$) & $2.73 \pm 0.11$ \\
{\rm Model} $0u$ ({\rm for} $T<3\;{\rm keV}$) & $0.35 \pm 0.16$ \\
{\rm Model} $0u$ ({\rm for} $T>3\;{\rm keV}$) & $2.61 \pm 0.14$ \\
{\rm Model} $1u$ ({\rm for} $T<2\;{\rm keV}$)& $0.05 \pm 0.07$ \\
{\rm Model} $1u$ ({\rm for} $T>2\;{\rm keV}$)& $2.71 \pm 0.11$ \\
{\rm Model} $1u$ ({\rm for} $T<3\;{\rm keV}$)& $0.26 \pm 0.13$ \\
{\rm Model} $1u$ ({\rm for} $T>3\;{\rm keV}$)& $2.62 \pm 0.14$ \\
\hline
Sample selection effects taken into account& \\
\hline
\citet[][]{Bharadwaj_2015} & $3.20 \pm 0.26$ \\
\citet[][]{Zou_2016} & $3.30 \pm 0.20$ \\
{\rm Model} $0$ ({\rm for} $T<2\;{\rm keV}$) & $2.55 \pm 0.55$ \\
{\rm Model} $0$ ({\rm for} $T>2\;{\rm keV}$) & $3.20 \pm 0.11$ \\
{\rm Model} $0$ ({\rm for} $T<3\;{\rm keV}$) & $2.05 \pm 0.29$ \\
{\rm Model} $0$ ({\rm for} $T>3\;{\rm keV}$) & $3.06 \pm 0.14$ \\
{\rm Model} $3$ ({\rm for all} $T$, $\beta_{\rm break}$) & $1.10 \pm 1.31$ \\
{\rm Model} $3$ ({\rm for all} $T$, $\beta$) & $3.27 \pm 0.12$ \\
{\rm Model} $1$ ({\rm for} $T<2\;{\rm keV}$) & $1.84 \pm 0.70$ \\
{\rm Model} $1$ ({\rm for} $T>2\;{\rm keV}$) & $3.20 \pm 0.11$ \\
{\rm Model} $1$ ({\rm for} $T<3\;{\rm keV}$) & $1.68 \pm 0.29$ \\
{\rm Model} $1$ ({\rm for} $T>3\;{\rm keV}$) & $3.08 \pm 0.14$ \\
{\rm Model} $5$ ({\rm for all} $T$, $\beta_{\rm break}$) & $1.19 \pm 1.17$ \\
{\rm Model} $5$ ({\rm for all} $T$, $\beta$) & $3.27 \pm 0.14$ \\
{\rm Model} $2$ ({\rm for} $T<2\;{\rm keV}$) & $2.52 \pm 0.60$ \\
{\rm Model} $2$ ({\rm for} $T>2\;{\rm keV}$) & $3.19 \pm 0.11$ \\
{\rm Model} $2$ ({\rm for} $T<3\;{\rm keV}$) & $2.06 \pm 0.30$ \\
{\rm Model} $2$ ({\rm for} $T>3\;{\rm keV}$) & $3.06 \pm 0.14$ \\
{\rm Model} $6$ ({\rm for all} $T$, $\beta_{\rm break}$) & $1.00 \pm 1.41$ \\
{\rm Model} $6$ ({\rm for all} $T$, $\beta$) & $3.26 \pm 0.14$ \\
{\rm Model} $4$ ({\rm for} $T<2\;{\rm keV}$) & $1.97 \pm 0.62$ \\
{\rm Model} $4$ ({\rm for} $T>2\;{\rm keV}$) & $3.18 \pm 0.11$ \\
{\rm Model} $4$ ({\rm for} $T<3\;{\rm keV}$) & $1.69 \pm 0.28$ \\
{\rm Model} $4$ ({\rm for} $T>3\;{\rm keV}$) & $3.07 \pm 0.15$ \\
{\rm Model} $7$ ({\rm for all} $T$, $\beta_{\rm break}$) & $1.20 \pm 1.23$ \\
{\rm Model} $7$ ({\rm for all} $T$, $\beta$) & $3.26 \pm 0.13$ \\
\hline
\end{tabular}
\end{table}

The separate modelling of scaling relations for groups and clusters implicitly assumes that these two types of objects constitute easily distinguishable populations, with minimum overlap with respect to their properties and/or clearly different formation and evolution paths. However, what we know about the process of large-scale structure formation and evolution suggests the same physical phenomena were responsible for the assembly of groups and clusters, though with varying relative influence across what would then constitute a single population. Consequently, scaling relations that jointly describe the dependencies between group and cluster properties should not show discontinuities. The simplest joint scaling relation model which ensures this just includes a change-point. It allows for a possible transition, in the way the properties being considered relate to each other, when moving from the group to the cluster regime, i.e from low to high temperature in the case of the  $L_{\rm X}-T$ relation. Interestingly, when this joint modelling is performed, by applying models 3 and 6 to the full sample, in contrast to applying models, respectively, 0 and 2, independently to the two temperature sub-samples, a widening of the difference between the $L_{\rm X}-T$ relation exponent in the group (low temperature) and cluster (high temperature) regimes is seen. All these models assume self-similar evolution for the normalisation of the $L_{\rm X}-T$ relation. When this assumption is relaxed, the same trend is found, although weaker. This can be seen by comparing the results of applying models 5 and 7 to the full sample, versus the results of applying models, respectively, 1 and 4 independently to the two temperature sub-samples. 

We also explored the possibility of the change in exponent of the $L_{\rm X}-T$ relation being driven by our prior assumptions that the scatter about the $L_{\rm X}-T$ relation does not change across the sample temperature range, and a possible change in exponent occurs abruptly. Allowing for the scatter to change abruptly at the same temperature as the exponent, thus introducing an extra free parameter with respect to model 7, does not lead to a significant change in the expected value for the exponent in the cluster regime. But it leaves the exponent in the group regime essentially unconstrained, although with an expected value even smaller than what was obtained assuming model 7. While allowing for a transition region with respect to both the exponent and scatter around the change-point, hence introducing yet another extra free parameter, the width of the transition region, we were unable to ascertain convergence for the MCMC produced by Gibbs sampling. This is probably due to a strong degeneracy between this parameter and one or more of the other parameters. Allowing for such a transition region should lead to a decrease in the expectation value for $\beta$, because systems with temperatures lower than that associated with the change-point would now also contribute to the estimation of the regression parameters for the higher temperature systems, and an increase in the expectation value for $\beta_{\rm break}$, due to the reciprocal effect. More data will be necessary to enable constraints to be imposed on the parameters of such complex models.

The self-similar expectation for the redshift evolution of the $L_{\rm X}-T$ relation is usually assumed to mean that its normalisation is proportional to $E(z)^{\gamma_{LT}}$ with $\gamma_{LT}=1$. Note that given our definition of $\gamma$, we have $\gamma_{LT}=\gamma+1$. However, \citet[][]{Maughan_2014} has also shown that the self-similar model implies more generally that
\begin{align}
\gamma_{LT}&=\gamma_{LM}-\frac{\beta_{LM}}{\beta_{TM}}\gamma_{TM}\\
&=\gamma_{LM}-\beta\,\gamma_{TM}
\label{gammaLT}
\end{align}

\noindent where $E(z)^{\gamma_{LM}}$ and $E(z)^{\gamma_{TM}}$ are assumed to describe the redshift evolution of the normalisation of the $L_{\rm X}-M$ and $T-M$ scaling relations, respectively. The self-similar expectation for $\gamma_{LT}$ is then recovered if $\gamma_{LM}$, $\beta$ and $\gamma_{TM}$ equal their self-similar values of $7/3$, $2$ and $2/3$, respectively. However, as discussed before, this will only hold true in the cluster regime, given that in the group regime the self-similar expectation for $\beta$ is $3/2$, and hence for $\gamma_{LT}$ is $4/3$. Further, if $\beta$ deviates from its self-similar expectation, as we have found seems to happen both in the group and cluster regimes, the same will happen for $\gamma_{LT}$ even assuming self-similar evolution for the $L_{\rm X}-M$ and $T-M$ relations. Given that our results suggest values for $\beta$ in the group and cluster regimes of $1.1\pm1.3$ and $3.3\pm0.1$, respectively, the self-similar expectation would then suggest values for $\gamma_{LT}$ in the group and cluster regimes of $1.6\pm0.9$ and $0.1\pm0.1$, respectively. But, if groups and clusters are assumed to belong to a single population of objects, distributed in the space of their properties in a continuous manner, then $\gamma_{LT}$ cannot change abruptly for any value of those properties, namely temperature or X-ray luminosity, unless the location of the respective change-point also changes (continuously) with redshift. Unfortunately, LIRA does not allow for this last option. Consistency with equation (\ref{eq8}), then implies that modelling the $L_{\rm X}-T$ relation using a change-point through LIRA in effect imposes different evolution rates with redshift for the normalisation of the $L_{\rm X}-M$ and/or $T-M$ scaling relations in the group versus the cluster regimes, and thus the location of the change-point for the differentially evolving scaling relation(s) among these to change with redshift. 

However, when models $1$ and $4$ are applied to the low temperature sub-samples the expected values for $\gamma_{LT}$ that are obtained are, respectively, $4.7\pm2.1$ and $4.4\pm2.1$ for $T<2\;{\rm keV}$, as well as $6.7\pm1.0$ for both models in the case of $T<3\;{\rm keV}$. When the same models are applied to the high temperature sub-samples those values are just, respectively, $0.3\pm0.5$ and $0.4\pm0.5$ for $T>2\;{\rm keV}$, as well as $-0.8\pm0.6$ and $-0.9\pm0.6$ for $T>3\;{\rm keV}$. This indicates that if a change-point is introduced in the modelling of the $L_{\rm X}-T$ relation, then its location should be allowed to change with redshift. In particular, a significantly faster evolving normalisation of the $L_{\rm X}-T$ relation in the group regime suggests the value for the temperature associated with its change-point is increasing with redshift. The inclusion of an extra parameter in LIRA to account for a possible evolution with redshift of the location of the change-point would thus be an interesting avenue to pursue, something we expect to do in future work.

The expected values obtained for $\gamma_{LT}$ with respect to the low temperature sub-samples are significantly higher than the self-similar expectation according to equation (\ref{eq8}), i.e. $1.6\pm0.9$. But statistically indistinguishable, with respect to $0.1\pm0.1$, for the high temperature sub-samples. A larger positive deviation with respect to the expected self-similar evolution for the $L_{\rm X}-T$ relation in the cluster regime, in accordance with equation (\ref{eq8}), was found by \citet[][]{Giles_2016}. On the other hand, \citet[][]{Reichert_2011} and \citet[][]{Clerc_2014}, as well as \citet[][]{Hilton_2012} using a distinct XCS sample, found negative evolution in the cluster $L_{\rm X}-T$ relation with respect to self-similar. However, the results of \citet[][]{Reichert_2011} and \citet[][]{Clerc_2014} could be explained by the difficulty of fully taking into account the distinct selection biases acting on the different cluster samples, spanning different redshift intervals, they used to determine the evolution of cluster properties, as extensively discussed in \citet[][]{Giles_2016} with respect to the results obtained by \citet[][]{Clerc_2014}. The reasons driving the differences between the results presented here and those described in \citet[][]{Hilton_2012} are discussed in the Appendix. The redshift evolution in the scatter of the $L_{\rm X}-T$ relation has been much less studied, but the results of \citet[][]{Mantz_2016} are in agreement with the negative evolution suggested by the data we used. 

The most recent numerical simulations, which include additional energy injection into the ICM/ICG from star formation and AGN feedback, as well as metal-dependent radiative cooling, are able to reproduce the observed $L_{\rm X}-T$ relation for clusters at low redshift, including a larger exponent than the self-similar expectation \citep[e.g.][]{Short_2013, Planelles_2014, Barnes_2017, LeBrun_2017,Truong_2018}. In agreement with our modelling assumption, \citet[][]{Truong_2018}, \citet[][]{Barnes_2017} and \citet[][]{LeBrun_2017} also found that, up to a redshift of 0.6, the exponent remained constant within the uncertainties. Further, their results for the evolution with redshift of the amplitude and scatter of the $L_{\rm X}-T$ relation, up to a redshift of 0.6, are in agreement with our results, supporting an increase in the amplitude above the self-similar expectation and a decrease in the scatter. Interestingly, \citet[][]{LeBrun_2017} also tried to model the $L_{\rm X}-T$ relation using a broken power-law, inserting a fixed change-point at $2$ keV. Contrary to our results they seem to find similar exponents for the $L_{\rm X}-T$ relation in the group and cluster regimes, although in agreement with our findings they see faster evolution in its amplitude in the group versus the cluster regime.

\section{Conclusions}
\label{s_conclusions}

We characterized the X-ray luminosity--temperature ($L_{\rm X}-T$) relation using a sample of 353 clusters and groups of galaxies with temperatures in excess of 1 keV and redshifts in the range $0.1 < z < 0.6$. We did not find strong statistical support for deviations from the usual modelling of the $L_{\rm X}-T$ relation as a single power-law, where the normalisation evolves self-similarly and the scatter remains constant with time. Nevertheless, assuming {\it a priori} the existence of the type of deviations considered, then faster evolution than the self-similar expectation for the normalisation of the $L_{\rm X}-T$ relation seems favoured, eventually signalling an increase with redshift in the average group/cluster X-ray luminosity at fixed temperature, as well as a decrease with redshift in the scatter about the $L_{\rm X}-T$ relation. Further, the preferred location for a possible temperature dependent change-point in the exponent of the $L_{\rm X}-T$ relation seems to be around 2 keV, possibly marking the transition between the group and cluster regimes. Our results also suggest an increase in the power-law exponent of the $L_{\rm X}-T$ relation when moving from the group to the cluster regime, and faster evolution in the former with respect to the later, driving the temperature dependent change-point towards higher values with redshift. We confirm past results that found an exponent for the $L_{\rm X}-T$ relation in the cluster regime significantly steeper than the self-similar expectation. But find the data suggests a shallower exponent with respect to it in the group regime. This implies that either the $L_{\rm X}-M$ relation is shallower, or the $T-M$ relation is steeper, in the case of groups and again with respect to the self-similar expectation.

The absence of statistical support for the existence of deviations from the standard modelling of the $L_{\rm X}-T$ cluster and group scaling relation as a single power-law, with a time-evolving normalization that follows the self-similar expectation and a non-evolving scatter, is somewhat of a surprise. In particular, because the self-similar approach actually predicts the $L_{\rm X}-T$ relation to deviate from a single power-law towards the lower temperature regime. This deviation could be weakened by non-gravitational effects, namely radiative cooling and AGN feedback, acting more strongly in groups versus clusters. However, if the impact of non-gravitational effects depends on the depth of the gravitational potential of the system they are acting on, then these effects should also leave an imprint on the redshift evolution of groups and clusters, because, at fixed mass, haloes are denser at higher redshifts. This stands in contrast with the absence of statistical support for the existence of deviations from the self-similar expectation for the redshift evolution of the $L_{\rm X}-T$ relation.

Hopefully, with the expected increased availability of large group and cluster samples, like those to be obtained from the Dark Energy Survey \citep[DES,][]{Abbott_2016} and eROSITA \citep[][]{Predehl_2014}, we will reach a stage where enough information will be available in the data to allow for the statistical detection of possible deviations from the standard modelling of the $L_{\rm X}-T$ relation. In particular, it would be important to expand the current group and cluster samples towards lower temperatures and higher redshift, in order to gain more leverage with respect to the exponent and evolution of the normalisation of the $L_{\rm X}-T$ relation in the group regime. In the meantime, the prior assumption of more complex models than the standard one does not seem warranted.

\section*{Acknowledgments}

This work was supported by Funda\c{c}\~{a}o para a Ci\^{e}ncia e Tecnologia (FCT) through national funds (UID/FIS/04434/2013) and by FEDER through COMPETE2020 (POCI-01-0145-FEDER-007672). L.E. acknowledges support through fellowship SFRH/BD/52138/2013, funded by FCT (Portugal) and POPH/FSE (EC). C.V.-C. acknowledges support from the Mexican National Council for Science and Technology scholarship scheme, grant number 411117. A.K.R. acknowledges support from the Science and Technology Facilities Council (grant number ST/P000252/1). S.B. acknowledges support from the UK Science and Technology Facilities Council via Research Training Grant ST/N504452/1. M.H. acknowledges financial support from the National Research Foundation, the South African Square Kilometre Array project, and the University of KwaZulu-Natal. J.M. acknowledges support from MPS, University of Sussex. M.S. acknowledges support by the Olle Engkvist Foundation (Stiftelsen Olle Engkvist Byggm\"astare).
 
\bibliographystyle{mnras}
\bibliography{refs}

\appendix
\section{Appendix}  

In this Appendix, we use the sample considered in \citet{Hilton_2012} to illustrate the impact on the estimation of temperatures and X-ray luminosities of the changes described in \citet{Manolopoulou_2018}, as well as on the characterization of the $L_{\rm X}-T$ relation.

\begin{figure*}
\begin{center}
\subfloat{\includegraphics[width=9.0cm]{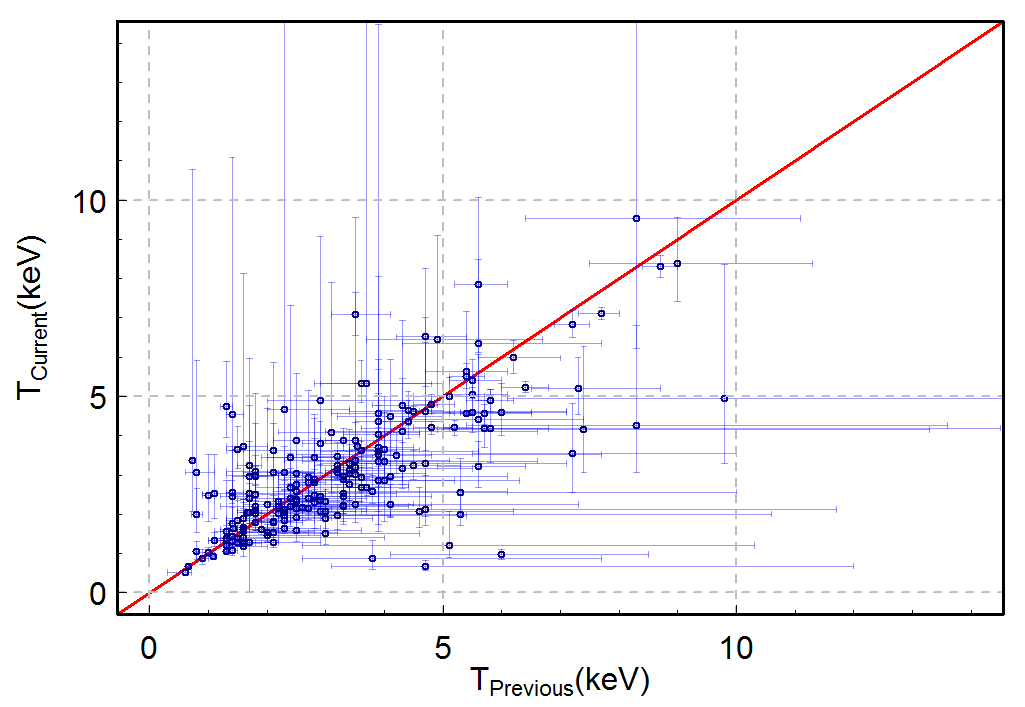}} 
\subfloat{\includegraphics[width=9.0cm]{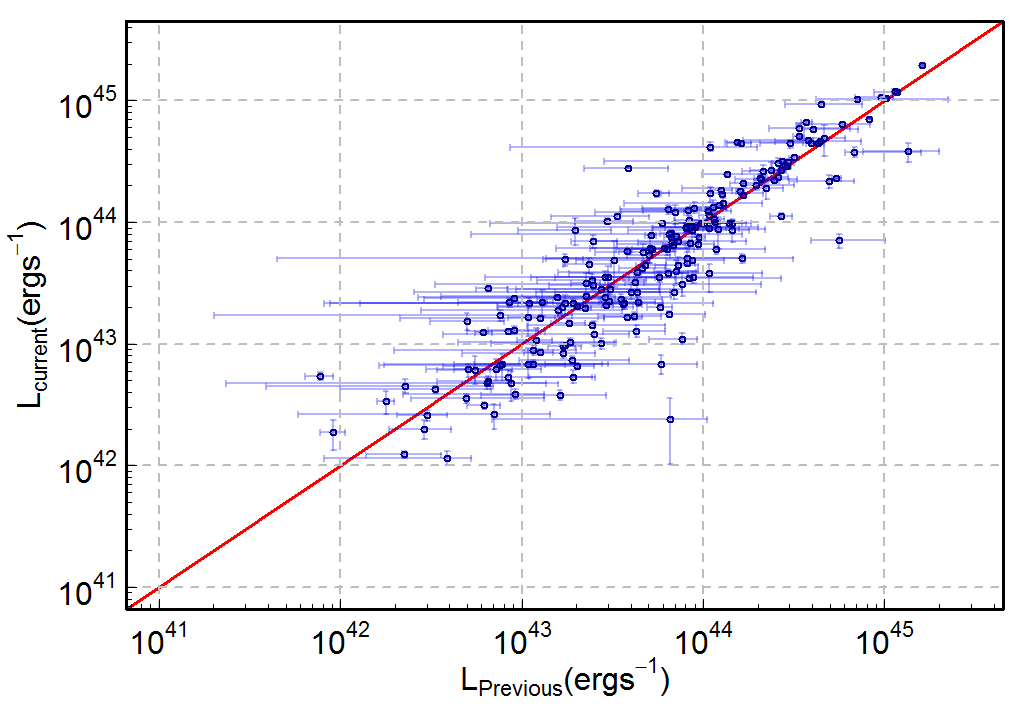}}
\end{center}
\caption{Comparison of the expected values and associated standard deviations for the temperatures and X-ray luminosities of 204 groups and clusters considered in \citet{Hilton_2012}, using the previous and current X-ray analysis methodology.}
\label{lt_comp}
\end{figure*}

In Fig. \ref{lt_comp}, we compare the expected values and associated standard deviations for the temperatures and X-ray luminosities for 204 out of the 211 groups and clusters considered in \citet{Hilton_2012}, as then estimated using the X-ray analysis methodology described in \citet{LloydDavies_2011} and using the current X-ray analysis methodology \citep{Manolopoulou_2018}. We omitted seven systems on the original sample from the comparison, because the current X-ray analysis methodology does not yield reliable values for the temperature and/or luminosity. The omission of these systems from the analysis should not have much of an impact, and thus henceforth we will often distinguish between the the original sample with 211 systems and its sub-sample of 204 systems. With respect to the X-ray luminosities, the current values for these systems are slightly higher on average than the previous ones, with the mean increasing from 1.40 to $1.46\times10^{44}$ erg s$^{-1}$, and have significantly smaller associated uncertainties. For the temperatures, on the contrary, the current values are on average lower than the previous ones, with the mean decreasing from 3.24 to 3.06 keV.

Although the model adopted for the $L_{\rm X}-T$ relation in \citet{Hilton_2012} is similar to what we here labelled as model 1u, we differ in the way the measurement uncertainties associated with the values of the sample temperatures and X-ray luminosities are propagated, and in how the likelihood is specified. With regards to the later, while \citet{Hilton_2012} penalized deviations between observed values and model predictions, for the temperatures and X-ray luminosities, using orthogonal and bisector metrics, here we focused only on such deviations with respect to the X-ray luminosity, at fixed temperature, in a manner analogous to what is commonly known as Y | X regression. If sample selection effects are not taken properly into account, these different approaches to likelihood specification can easily lead to significantly different results. Applying our model 1u to the sample in \citet{Hilton_2012}, assuming the expected values and associated standard deviations for the observed temperatures and X-ray luminosities as obtained with the previous X-ray analysis methodology, we obtain for the expected values and standard deviations: $\alpha=44.22\pm0.07$, $\beta=2.21\pm0.13$, $\sigma=0.28\pm0.02$ and $\gamma=0.11\pm0.46$. For comparison, the bisector method yielded $\alpha=44.41\pm0.04$, $\beta=2.65\pm0.09$, $\sigma=0.34\pm0.02$ and $\gamma=-0.7\pm0.3$, while applying the orthogonal method resulted in $\alpha=44.63\pm0.07$, $\beta=3.02\pm0.15$, $\sigma=0.27\pm0.03$ and $\gamma=-1.9\pm0.5$ \citep[Table 3,][]{Hilton_2012}. Therefore, as was somewhat expected given the different methodologies used, our analysis of the same data considered in \citet{Hilton_2012} leads to markedly different results. However, it should be noted that our results differ as much from those yielded by the bisector method, as these differ from those that arise from applying the orthogonal method.

\begin{figure*}
\includegraphics[width=\columnwidth]{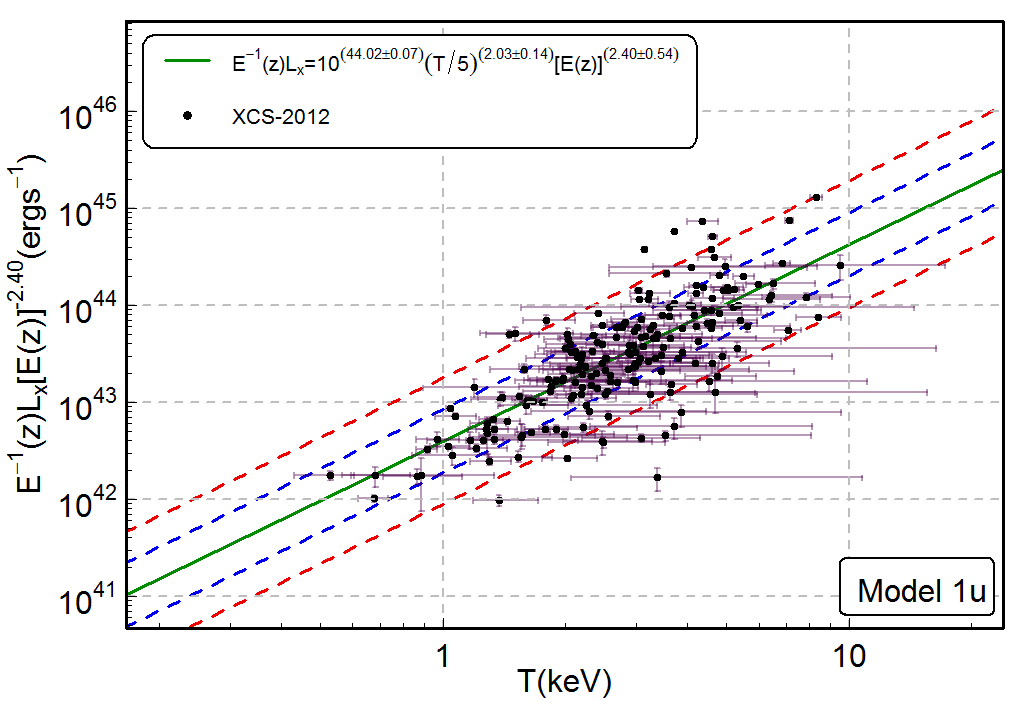}
\includegraphics[width=\columnwidth]{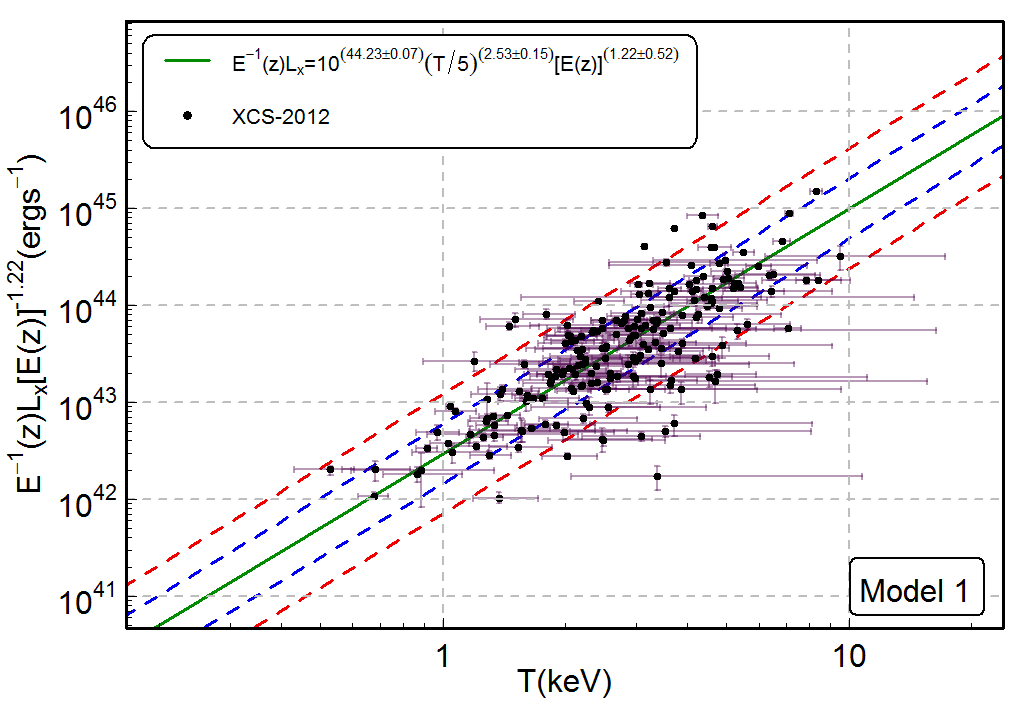}
\caption{The \lxt relation assuming model 1, without (left, model 1u) and with (right, model 1) sample selection effects taken into account. Each point indicates the temperature, $T$, in keV, and re-scaled bolometric X-ray luminosity, $ L_{\rm X}$, expected for each of the 204 systems in the \citet{Hilton_2012} sample considered here, given the X-ray data. The re-scaling of $L_{\rm X}$ with the cluster redshift, $z$, is performed dividing $L_{\rm X}$ by $E(z)^{1+\overline\gamma}$, where $\overline\gamma$ is the expected (i.e. mean) value for $\gamma$. The error bars associated with each point identify the $1\sigma$ uncertainty intervals. The solid regression line was determined by fixing the model hyper-parameters to their expected (i.e. mean) values. The amplitude of the intrinsic vertical scatter about the regression line is indicated by the dashed lines ($1\sigma$, inner blue; $2\sigma$, outer red).}
\label{lt_2012}
\end{figure*}

We have also applied models 1u and 1 to the sample in \citet{Hilton_2012}, but now assuming the expected values and associated standard deviations for the observed temperatures and X-ray luminosities as obtained with the current X-ray analysis methodology. Note that, in the case of model 1, as before we take into account the XCS selection function, but now we do not assume {\it a priori} that a lower threshold in temperature as been imposed in the process of sample selection. In Table \ref{table1a} we present the results, and in Fig. \ref{lt_2012} we plot the distribution of the data points associated with the \citet{Hilton_2012} sample, as well as the regression lines obtained assuming the expected (i.e. mean) values for the relevant model hyper-parameters, either assuming model 1u or 1. As happened when the same models were applied to the XCS-SDSS data, we find an increase in the expected values for $\alpha$ and $\beta$, as well as a decrease in the expected value for $\gamma$, when sample selection effects are taken into account.

\begin{table}
\caption{The mean and standard deviation for the marginal posterior distributions of the interesting hyper-parameters associated with each model, given 204 groups and clusters in the \citet{Hilton_2012} sample. The expected values and associated standard deviations for the observed temperatures and X-ray luminosities were obtained with the current X-ray analysis methodology.}
\label{table1a}
\def\arraystretch{1.5}
\begin{tabular}{|c|c|c|c|c|}
\hline
Model & $\alpha$ & $\beta$ & $\sigma$ & $\gamma$\\
\hline
$1{\rm u}$ & $44.02, 0.07$ & $2.03, 0.14$ & $0.33, 0.02$ & $2.40, 0.54$\\
$1$ & $44.23, 0.07$ & $2.53, 0.15$ & $0.31, 0.02$ & $1.22, 0.52$\\
\hline
\end{tabular}
\end{table}

The results obtained by applying model 1u to the sample in \citet{Hilton_2012}, assuming either the expected values and associated standard deviations for the observed temperatures and X-ray luminosities as obtained with the previous X-ray analysis methodology, or those obtained with the current X-ray analysis methodology, are compatible within the statistical uncertainties, except for the evolution parameter $\gamma$. The previous X-ray data suggested much weaker evolution with redshift for the normalisation of the $L_{\rm X}-T$ relation than the current X-ray data. On the other hand, the expected values for the model parameters $\alpha$, $\beta$ and $\gamma$, are significantly different from what was obtained assuming the XCS-SDSS data, for both models 1u and 1. This would be expected in the former case, given the absence of correction for the different selection procedures behind the assembly of both samples. But it is somewhat surprising in the later case, although the discrepancy becomes smaller. Possibly it reflects the greater difficulty of modelling the more complex nature of the selection function associated with the assembly of the sample used in \citet{Hilton_2012}.

\label{lastpage}

\end{document}